# Chirality Transfer to the Magnetic Sublattice in the Hybrid Perovskite (R)-/(S)-3-Fluoropyrrolidinium Copper(II) Chloride

Zheng Zhang, Mingyu Xu, Jose L. Gonzalez Jimenez, Stephen Zhang, Weiwei Xie, Xianghan Xu, and Daniel B. Straus*

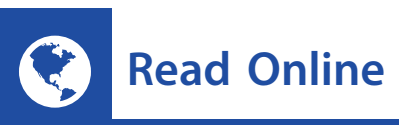

Read Online

ACCESS | Metrics & More | Article Recommendations | Supporting Information

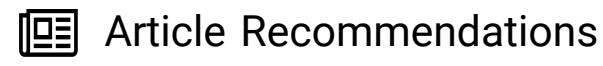

**ABSTRACT:** Incorporating chiral organic cations into organic−inorganic hybrid materials has been shown to enable the inorganic sublattice to display chiroptical properties. We report a new two-dimensional magnetic ($S = 1/2$) chiral metal halide perovskite, (R)- and (S)-$(C_4H_9FN)_2CuCl_4$ (where $(C_4H_9FN)^+$ is 3-fluoropyrrolidinium), which consists of Cu−Cl inorganic layers separated by $(C_4H_9FN)^+$ organic cations. The presence of the chiral $(C_4H_9FN)^+$ organic cation induces the formation of chiral magnetic order, even though the inorganic sublattice itself is nearly structurally centrosymmetric. We also report the racemic variant, containing an equal amount of (R)- and (S)- cations, which shows no evidence of chiral magnetic order. When the magnetic susceptibility is measured perpendicular to the inorganic Cu−Cl layer propagation direction, an antiferromagnetic phase transition at Néel temperature $T_N$ = 2.23 K is observed in both the chiral and racemic materials, and the existence of the magnetic phase transition is supported by specific heat capacity measurements. Field-induced magnetic chirality is observed through the existence of a second-order magnetoelectric effect in the chiral variant, while no magnetoelectric signal is observed for the racemic material, indicating the absence of magnetic chirality. Our findings demonstrate that materials exhibiting chiral magnetic order can be created through the incorporation of a chiral cation into an organic−inorganic hybrid magnetic material, potentially allowing for the design of tailored materials that combine chiral magnetism with other desirable optical and electronic properties that come from structural chirality.

(R)/(S)-3-Fluoropyrrolidinium Copper(II) Chloride

Chiral Organic Cation

Magnetic Chirality

## 1. INTRODUCTION

Chiral magnetic semiconductors that demonstrate fascinating chiral, semiconducting, and magnetic properties are highly sought after for practical applications such as new forms of high-speed nonvolatile memory and chiral magnetic field sensors.[1−4] For a material to exhibit magnetic chirality, it must have chiral spin-ordered states and/or chiral spin textures.[5,6] Structurally chiral magnetic materials are not guaranteed to exhibit chiral magnetic ground states, with an example being $Ni_3TeO_6$.[5,7] Magnetic chirality has been observed in achiral materials including $YMn_6Sn_6$,[8] further demonstrating that structural and magnetic chirality are not always connected.[8−11] In the structurally chiral inorganic materials $Ba_3NbFe_3Si_2O_{14}$,[12] $Cr_{1/3}TaS_2$,[13] and doped $Ni_3TeO_6$,[14,15] magnetic chirality is observed, even though the sublattice of magnetic atoms is structurally achiral. There are also materials such as $CsCuCl_3$,[16] MnSi,[17] and $CrNb_3S_6$,[18] where the materials are structurally chiral, the magnetic sublattices are also structurally chiral, and magnetic chirality has been observed. Furthermore, even in materials that exhibit both structural and magnetic chirality, the fundamental relation between the two is unclear. Some evidence suggests that in materials that are both structurally and magnetically chiral, the handedness of the magnetic structure matches that of the crystal lattice.[19,20] However, it has also been reported that if a structurally and magnetically chiral material is doped, the handedness of the chiral magnetic sublattice will flip and become opposite to the material's structural handedness.[21−24]

The handedness of chiral all-inorganic materials is very difficult to control because in typical crystal growth experiments, an equal amount of left- and right-handed crystals will form as the energies of formation for left- and right-handed materials are equal.[25] However, incorporating a chiral organic molecule of a single handedness into a material causes the composite organic−inorganic hybrid material to be chiral, maintaining the handedness of the chiral molecule.[26,27] Until recently, organic−inorganic chiral hybrid magnets were limited to molecular magnets, where chiral ligands are directly bound to the metals,[28−31] but advances in the synthesis of chiral organic−inorganic hybrid materials, such as perovskites, have allowed new classes of chiral magnetic semiconductors to be synthesized. The first chiral organic−inorganic hybrid halide perovskites were first synthesized in 2003.[27] In almost all hybrid perovskites, there are no traditional chemical bonds between the organic and inorganic parts of the material; the interactions are electrostatic. In 2016, it was demonstrated that in organic−inorganic hybrid perovskites containing a chiral organic cation, chirality is transferred from the organic to the inorganic lattice despite the lack of chemical bonds between the organic cations and the inorganic framework, and the transfer of chirality was identified by the presence of natural circular dichroism in optical transitions attributed to the material's inorganic framework.[32] The mechanism of chirality transfer to the inorganic framework is thought to originate from asymmetric hydrogen bonding interactions inducing chiral distortions in the inorganic framework[33,34] or from asymmetric orbital polarization induced by the chiral organic cations.[35]

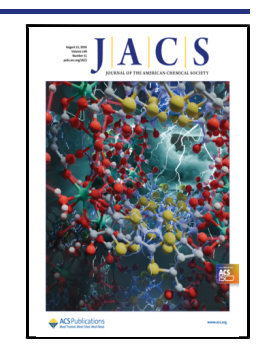

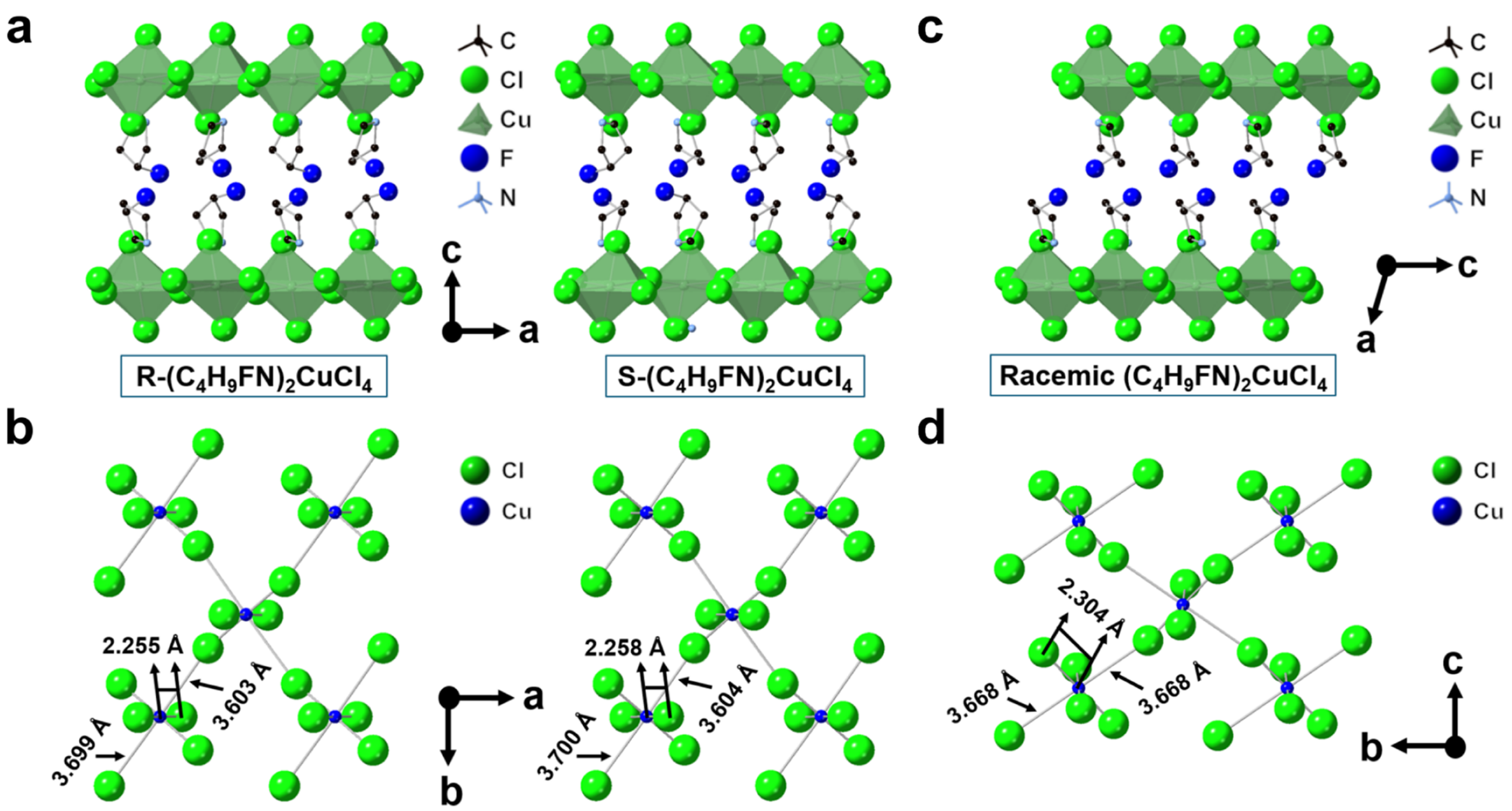

**Figure 1.** Crystal structures of (a) and (b) (R)- and (S)-$(C_4H_9FN)_2CuCl_4$, and (c) and (d) racemic $(C_4H_9FN)_2CuCl_4$. Hydrogen atoms are omitted for clarity.

In centrosymmetric materials, the magnetic behavior of a material is determined by the sign of the exchange interaction between electrons on interacting open shell atoms. If only nearest-neighbor magnetic interactions are considered, the classical Heisenberg Hamiltonian can be used: $\hat{H} = -J\sum_{\langle i,j\rangle}\hat{S}_i\cdot\hat{S}_j$, where $J$ is the exchange coupling constant between nearest-neighbor lattice points $i$ and $j$, and $\hat{S}$ is a unit vector that points in the direction of the spin.[36−38] Under this convention, an interaction is ferromagnetic if $J$ is positive and antiferromagnetic if $J$ is negative.

In chiral magnets, an antisymmetric contribution to the Hamiltonian must be considered, which leads to an asymmetric electronic structure. For spins $\hat{S}_i$ and $\hat{S}_j$, $\hat{H}_{i,j} = \overrightarrow{D_{ij}}\cdot(\hat{S}_i\times\hat{S}_j)$. This is also called the Dzyaloshinskii−Moriya (DM) interaction.[39,40] The DM interaction in chiral materials can lead to the formation of helical magnetic spin vortices.[41] Other chiral magnetic structures can also develop, such as a helical chiral magnetic structure, which arises from the coupling between the DM interaction and a ferromagnetic Heisenberg interaction.[19,39]

The first structurally chiral organic−inorganic hybrid magnetic perovskites were synthesized in 2020, which were the ferromagnetic two-dimensional (2D) materials (R)- and (S)-$(MPEA)_2CuCl_4$ (where MPEA is $\beta$-methylphenethylamine).[42] Since then, additional chiral hybrid materials have been reported including the antiferromagnetic (R)-$(MPEA)_4AgFeCl_8$,[43] antiferromagnetic (R)- and (S)-$(MPEA)_2MnCl_4(H_2O)$,[44] and the multiferroic (R)- and (S)-$(MPEA)_2CuCl_4$ exhibiting A-type (or easy-axis)[45] antiferromagnetism and intralayer ferroelectricity.[46] However, magnetic chirality (i.e., a spin arrangement with no mirror symmetry that does not change handedness under time reversal)[5] has not been experimentally demonstrated in any organic−inorganic chiral magnetic material.

In this work, we synthesized the new chiral magnetic perovskite (R)- and (S)-$(C_4H_9FN)_2CuCl_4$ (where $(C_4H_9FN)^+$ is (R)- or (S)-3-fluoropyrrolidinium) and demonstrate that its magnetic structure exhibits magnetic chirality even though the inorganic Cu−Cl sublattice is nearly centrosymmetric, as within reasonable tolerances, it exhibits inversion symmetry in both (R)- and (S)-$(C_4H_9FN)_2CuCl_4$. In contrast, racemic $(C_4H_9FN)_2CuCl_4$, containing an equal amount of (R) and (S) cations, does not exhibit magnetic chirality, which is expected because the racemic variant is centrosymmetric. Structurally, (R)-, (S)-, and racemic $(C_4H_9FN)_2CuCl_4$ are highly distorted two-dimensional (2D) Ruddlesden−Popper phase organic−inorganic hybrid perovskites, where the metal-halide (Cu−Cl) 2D layers or slabs are separated by $(C_4H_9FN)^+$ organic cations. (R)- and (S)-$(C_4H_9FN)_2CuCl_4$ are optically active and have mirrored circular dichroism (CD) spectra, as expected for enantiomorphic chiral materials. Magnetically, (R)-, (S)-, and racemic $(C_4H_9FN)_2CuCl_4$ are A-type or easy-axis antiferromagnets, where ferromagnetic interactions dominate within each 2D Cu−Cl layer, but adjacent layers order antiferromagnetically. When the magnetic susceptibility is measured with the field applied perpendicular to the inorganic Cu−Cl layer propagation direction, we observe a transition at the Néel temperature $T_N$ = 2.23 K in all three materials, with a corresponding phase transition observed in specific heat capacity measurements. We also perform AC magnetic susceptibility measurements and find no frequency dependence for (R)- and (S)-$(C_4H_9FN)_2CuCl_4$ both above and below the Néel temperature, while racemic $(C_4H_9FN)_2CuCl_4$ shows a strong frequency dependence below $T_N$. Most importantly, the presence of the magnetoelectric effect in (R)- and (S)-$(C_4H_9FN)_2CuCl_4$ demonstrates the presence of magnetic chirality, while no magnetoelectric signal is observed for racemic $(C_4H_9FN)_2CuCl_4$, indicating an absence of magnetic chirality. Our findings demonstrate that chiral organic cations induce magnetic chirality in the inorganic (magnetic) sublattice, enabling the synthesis of enantiopure samples of magnetically chiral crystals where the structural handedness is defined by the chiral organic cation. Magnetically chiral organic−inorganic hybrid materials including (R)- and (S)-$(C_4H_9FN)_2CuCl_4$ can thus be considered for new classes of spintronic devices such as high-speed nonvolatile memory[1,2] and chiral magnetic field sensors,[4] as well as provide insight into potential relations between the DM interaction, magnetoelectricity, and the chiral-induced spin selectivity (CISS) effect.[47−49]

**Table 1. Crystallographic Collection and Structural Refinement Parameters**

| Structure | (R)-$(C_4H_9FN)_2CuCl_4$ | (S)-$(C_4H_9FN)_2CuCl_4$ | Racemic $(C_4H_9FN)_2CuCl_4$ |
|---|---|---|---|
| Formula weight (g/mol) | 385.58 | 385.58 | 385.58 |
| Temperature (K) | 300 | 300 | 300 |
| Wavelength (Å) | 0.71073 (Mo k$\alpha$) | 0.71073 (Mo k$\alpha$) | 1.54178 (Cu k$\alpha$) |
| Crystal system | Orthorhombic | Orthorhombic | Monoclinic |
| Space group | $P2_12_12_1$ (#19) | $P2_12_12_1$ (#19) | $P2_1/c$ (#14) |
| $Z$ | 4 | 4 | 2 |
| Unit cell parameters | $a$ = 7.7559(4) Å | $a$ = 7.7598(3) Å | $a$ = 10.9445(2) Å |
| | $b$ = 8.7973(4) Å | $b$ = 8.8006(3) Å | $b$ = 9.0521(2) Å |
| | $c$ = 21.1758(10) Å | $c$ = 21.1880(6) Å | $c$ = 7.56760(10) Å |
| | | | $\beta$ = 105.4050(10)° |
| Volume (Å$^3$) | 1444.85(12) | 1446.95(9) | 722.79(2) |
| Density (g/cm$^3$) | 1.773 | 1.770 | 1.772 |
| Absorption coefficient ($\mu$) (mm-1) | 2.253 | 2.250 | 8.994 |
| $\theta_{min}$ - $\theta_{max}$ (°) | 5.014 to 61.102 | 5.012 to 61.218 | 12.886 to 144.562 |
| Reflections collected | 27540 | 29572 | 16638 |
| Independent reflections | 4428 | 4446 | 1428 |
| $Ra$ indices ($I > 2\sigma(I)$) | $R_1$ = 0.0402 | $R_1$ = 0.0367 | $R_1$ = 0.0214 |
| | $wR_2$ = 0.0842 | $wR_2$ = 0.0903 | $wR_2$ = 0.0604 |
| Goodness-of-fit on $F2$ | 1.048 | 1.017 | 1.063 |
| Largest diff. peak/hole (e-/Å3) | 0.63/−0.60 | 0.53/−0.56 | 0.31/−0.26 |
| Flack parameter | 0.009(15) | 0.027(11) | - |

## 2. RESULTS AND DISCUSSION

### 2.1. Structural Characterization

Figures 1 and S1 depict the crystal structures of (R)-, (S)-, and racemic $(C_4H_9FN)_2CuCl_4$. These materials are structurally similar to the Pd analogs that we recently reported[50] and are highly distorted 2D Ruddlesden−Popper phase hybrid perovskites, where layers of corner-sharing $CuCl_6$ octahedra are separated from one another by $(C_4H_9FN)^+$ cations (Figure 1). Within a $CuCl_6$ octahedron, there are four short Cu−Cl bonds that are 2.26 Å in (R)-/(S)-$(C_4H_9FN)_2CuCl_4$ and 2.30 Å in racemic $(C_4H_9FN)_2CuCl_4$, as well as two long (3.6−3.7 Å) Cu−Cl contacts (Figure 1b and 1d). The observed short Cu−Cl bond lengths are in good agreement with the Cu−Cl bond lengths observed in other Cu(II)-based metal halides such as $CsCuCl_3$ and $(PMA)_2CuCl_4$ (where PMA is phenylmethylammonium).[51] The elongated Cu−Cl bond contact of approximately 3.60−3.70 Å is due to Jahn−Teller distortions, as noted in other hybrid (organic−inorganic) Cu(II) halides including $MA_2CuCl_4$ (where MA is methylammonium) and $(PMA)_2CuCl_4$.[51,52] In (R)-/(S)-$(C_4H_9FN)_2CuCl_4$, the two Cu−Cl long contacts have a noticeable difference in length (∼0.1 Å), while the two long contacts are equivalent in length in the racemic analog. Moreover, we observe no significant difference in physical density between the chiral and racemic $(C_4H_9FN)_2CuCl_4$ variants (Table 1). (R)-/(S)-$(C_4H_9FN)_2CuCl_4$ crystallize in the orthorhombic $P2_12_12_1$ (#19) space group, which is in higher symmetry than their Pd analogs, which crystallize in the monoclinic $P2_1$ space group.[50] The racemic $(C_4H_9FN)_2CuCl_4$ grows in the monoclinic $P2_1/c$ (#14) space group and is isostructural to its Pd analog. Both (R)-/(S)- and racemic $(C_4H_9FN)_2CuCl_4$ crystals are dark green in color. Detailed single-crystal data and structural refinement parameters are shown in Table 1. Importantly, the inorganic Cu−Cl lattice is nearly centrosymmetric in both (R)- and (S)-$(C_4H_9FN)_2CuCl_4$. The greatest deviation from true inversion symmetry (0.193 Å and 0.195 Å in the (R)- and (S)-variants, respectively) is less than the radius of the 50% probability thermal ellipsoid at room temperature, which was determined using the ADDSYM EXACT function in PLATON (Supporting Information); the centrosymmetric inorganic sublattice output by PLATON is compared with the experimental structure in Figure S2.[53]

Figure S3 shows the measured experimental powder X-ray diffraction (PXRD) pattern of $(C_4H_9FN)_2CuCl_4$ compared to the calculated pattern. The experimental PXRD pattern shows good agreement with the calculated pattern given by single-crystal XRD measurements, indicating the sample that we used for PXRD analysis is phase-pure. To check the stability of $(C_4H_9FN)_2CuCl_4$ against ambient air exposure, we performed periodic PXRD measurements over time to monitor changes in the PXRD pattern (see Figure S4). Based on the obtained PXRD data, both (R)- and (S)-$(C_4H_9FN)_2CuCl_4$ show a decrease in peak intensity after 4 weeks of exposure to air, suggesting reduced crystallinity but no transition to a new crystalline phase. Moreover, after 4 weeks, we observed a color change on the powder sample's surface, implying $(C_4H_9FN)_2CuCl_4$ is not completely stable in air, though the degradation occurs over a time frame of weeks to months. Degradation of single crystals is likely insignificant compared to powder due to the much smaller surface area-to-volume ratio.

### 2.2. Optical Characterization

Figure 2 shows the obtained ultraviolet−visible (UV−vis) absorbance and circular dichroism (CD) data using the deposited chiral and racemic $(C_4H_9FN)_2CuCl_4$ films. X-ray diffraction patterns of the (R)-/(S)-$(C_4H_9FN)_2CuCl_4$ and racemic $(C_4H_9FN)_2CuCl_4$ thin films are shown in Figure S5. CD spectra are corrected for linear dichroism and linear birefringence effects (Figure S6).[54−56] The UV−vis absorbance data show two distinct absorption features: a peak centered at 380 nm and a shoulder at ∼450 nm. Based on the CD spectra, (R)- and (S)-$(C_4H_9FN)_2CuCl_4$ thin films are optically active, indicating that the organic cation transfers chirality into the inorganic matrix. As expected, the CD spectra for (R)- and (S)-$(C_4H_9FN)_2CuCl_4$ exhibit features of opposite signs, with maxima/minima at 383 and 353 nm with the spectra intersecting at 367 nm. The CD sign inversion at 367 nm is an indication of the Cotton effect, which is expected to be observed near a feature in the absorption spectrum.[56] As expected, a negligible CD signal was observed for racemic $(C_4H_9FN)_2CuCl_4$. This observation is similar to the Pd analog that we reported,[50] where (R)-/(S)-$(C_4H_9FN)_2PdCl_4$ are optically active, while we observe no CD signal from (R)-/(S)-$(C_5H_{12}N)_2PdCl_4$ ($(C_5H_{12}N)^+$ is 3-methylpyrrolidinium), highlighting the importance of the chiral organic cation and the interaction between the chiral cation and the inorganic framework in determining the chiroptical properties of the material. Furthermore, the calculated absolute values of the anisotropy factor $g_{CD} = \frac{\text{CD (mdeg)}}{32980 \times \text{A}}$, where A is the absorbance, for (R)- and (S)-$(C_4H_9FN)_2CuCl_4$ at 415 nm are $0.63 \times 10^{-4}$ and $1.44 \times 10^{-4}$ (Figure S7), respectively. These values are similar to those of (R)-/(S)-$(THBTD)_2CuCl_6$ ($2.8 \times 10^{-4}$ for (R), $6.8 \times 10^{-4}$ for (S), where THBTD is 4,5,6,7-tetrahydro-benzothiazole-2,6-diamine), and an order of magnitude lower than those of (R)-/(S)-$(1,2\text{-NEA})_2CuCl_4$ ($\sim 2.9 \times 10^{-3}$, where 1,2-NEA is 1-(2-naphthyl)ethylammonium).[57,58]

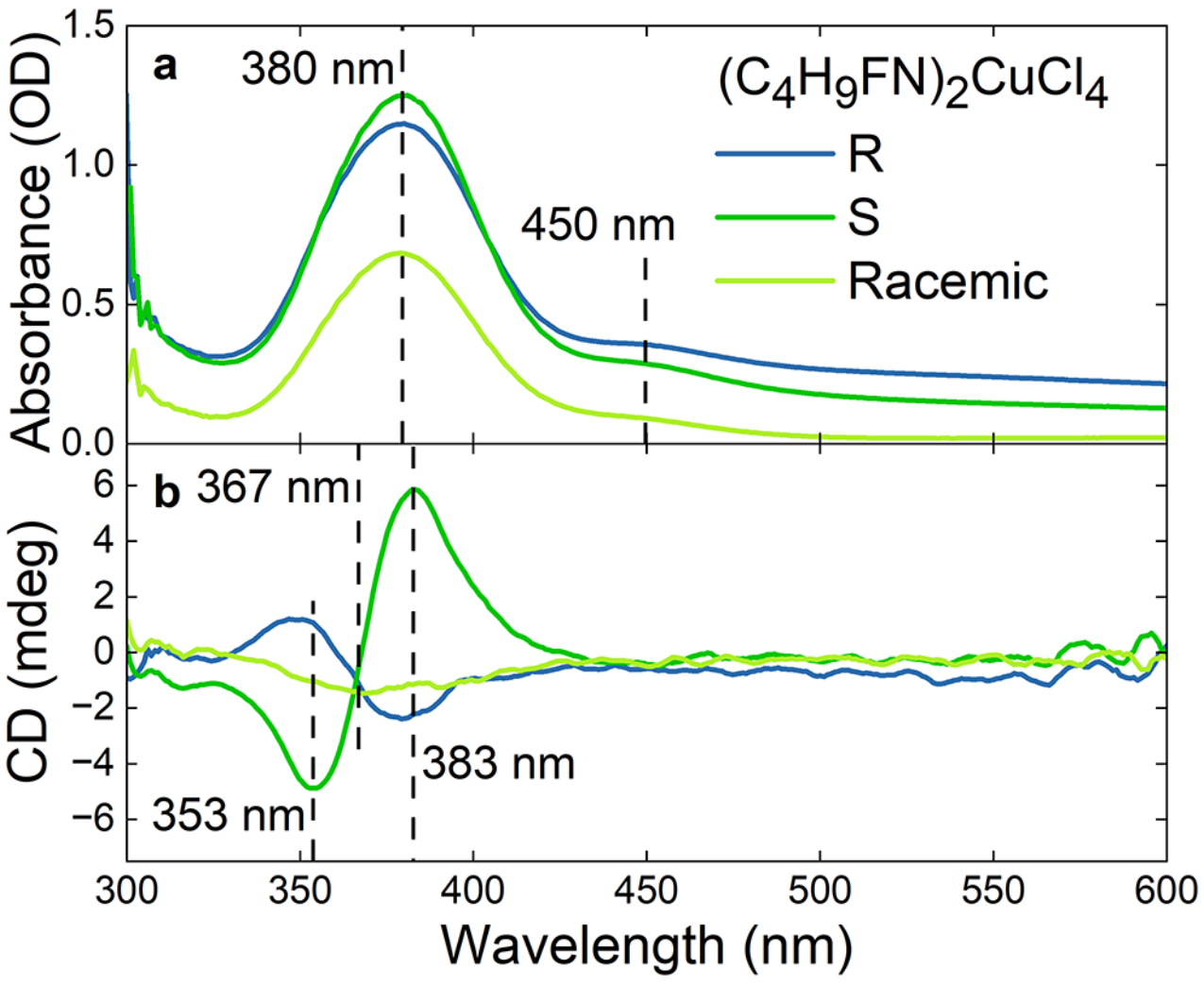

Figure 2. (a) UV−vis absorption and (b) CD spectra of (R)-, (S)-, and racemic $(C_4H_9FN)_2CuCl_4$ thin films.

### 2.3. Magnetic Characterization

Figure 3 shows the zero-field-cooled inverse magnetic susceptibility vs temperature for powdered (R)-, (S)-, and racemic $(C_4H_9FN)_2CuCl_4$ at an applied field of 1000 Oe. At temperatures far from any phase transitions, the inverse susceptibility $1/\chi$ should be linear with temperature following the Curie−Weiss law $\frac{1}{(\chi - \chi_0)} = \left(\frac{1}{C}\right)T - \left(\frac{\theta_{CW}}{C}\right)$, where $\chi$ is the susceptibility, $\chi_0$ is the temperature-independent contribution to the susceptibility that likely comes from core diamagnetism,[59] C is the Curie constant, and $\theta_{CW}$ is the Weiss temperature; the x-intercept of the extrapolated linear fit is $\theta_{CW}$. This relation holds true for our data: the inverse susceptibility from 50−300 K is well-fit by the Curie−Weiss law (all fits have adjusted $R^2$ values greater than 0.99), and the fits are shown as solid lines in Figure 3 with dashed lines extrapolating the linear fit to $\theta_{CW}$, and fit parameters are listed in Table 2. The effective moment was calculated using the equation $\mu_{eff} = \sqrt{\frac{3k_B}{N_A\mu_B^2}C}$, expressed in Bohr magnetons ($\mu_B$). The positive Weiss temperatures $\theta_{CW}$, which can be observed in Figure 3 by the positive value of the *x*-intercept, indicate that ferromagnetic interactions dominate. For all variants, the effective moment is greater than the spin-only moment for an $S = 1/2$ system ($\mu = 1.73\mu_B$), which is commonly observed in Cu(II)-containing materials and results from spin−orbit coupling interactions.[60,61] Figure S8 shows both zero-field-cooled and field-cooled susceptibility measurements on the chiral and racemic variants. These traces overlap perfectly, making the presence of bulk ferromagnetism unlikely despite the positive Weiss temperatures.[62]

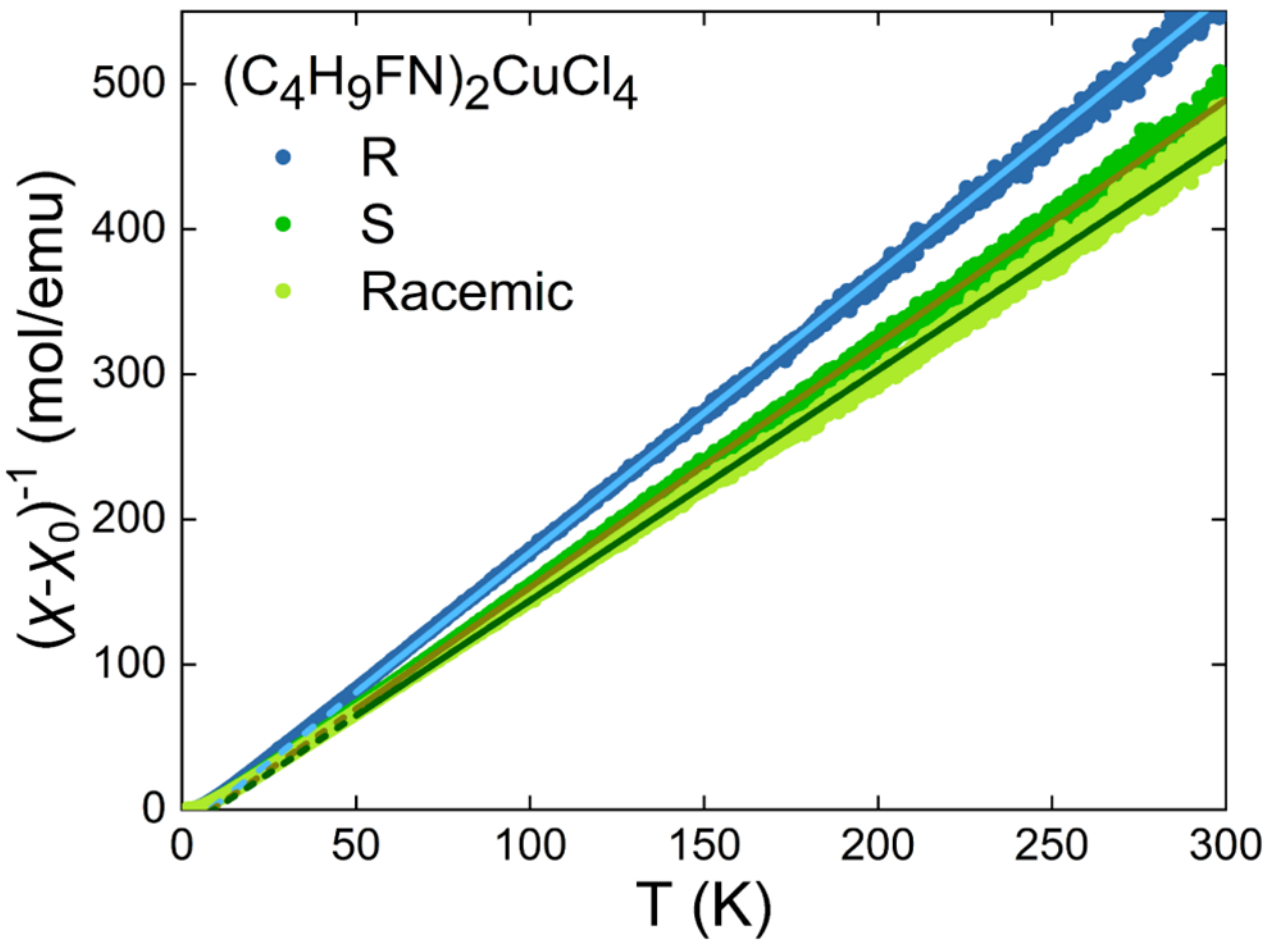

Figure 3. Plot of the inverse susceptibility $(\chi - \chi_0)^{-1}$ versus temperature for powder samples of (R)-, (S)-, and racemic $(C_4H_9FN)_2CuCl_4$. Curie−Weiss fits are shown as solid lines with extrapolations of the fits as dashed lines.

**Table 2. Parameters Obtained from Curie−Weiss Fits of the Magnetic Susceptibility Data for $(C_4H_9FN)_2CuCl_4$ Powder**

| | (R) | (S) | Racemic |
|---|---|---|---|
| $\mu_{eff}$ ($\mu_B$) | 2.04 | 2.18 | 2.24 |
| $\theta_{CW}$ (K) | 8.07 | 8.69 | 9.23 |
| $\chi_0$ (emu/mol) | −0.00131 | −0.00109 | −0.00164 |

We further carried out magnetic characterizations for the synthesized chiral and racemic $(C_4H_9FN)_2CuCl_4$ single crystals; field-cooled measurements are shown in Figure 4.

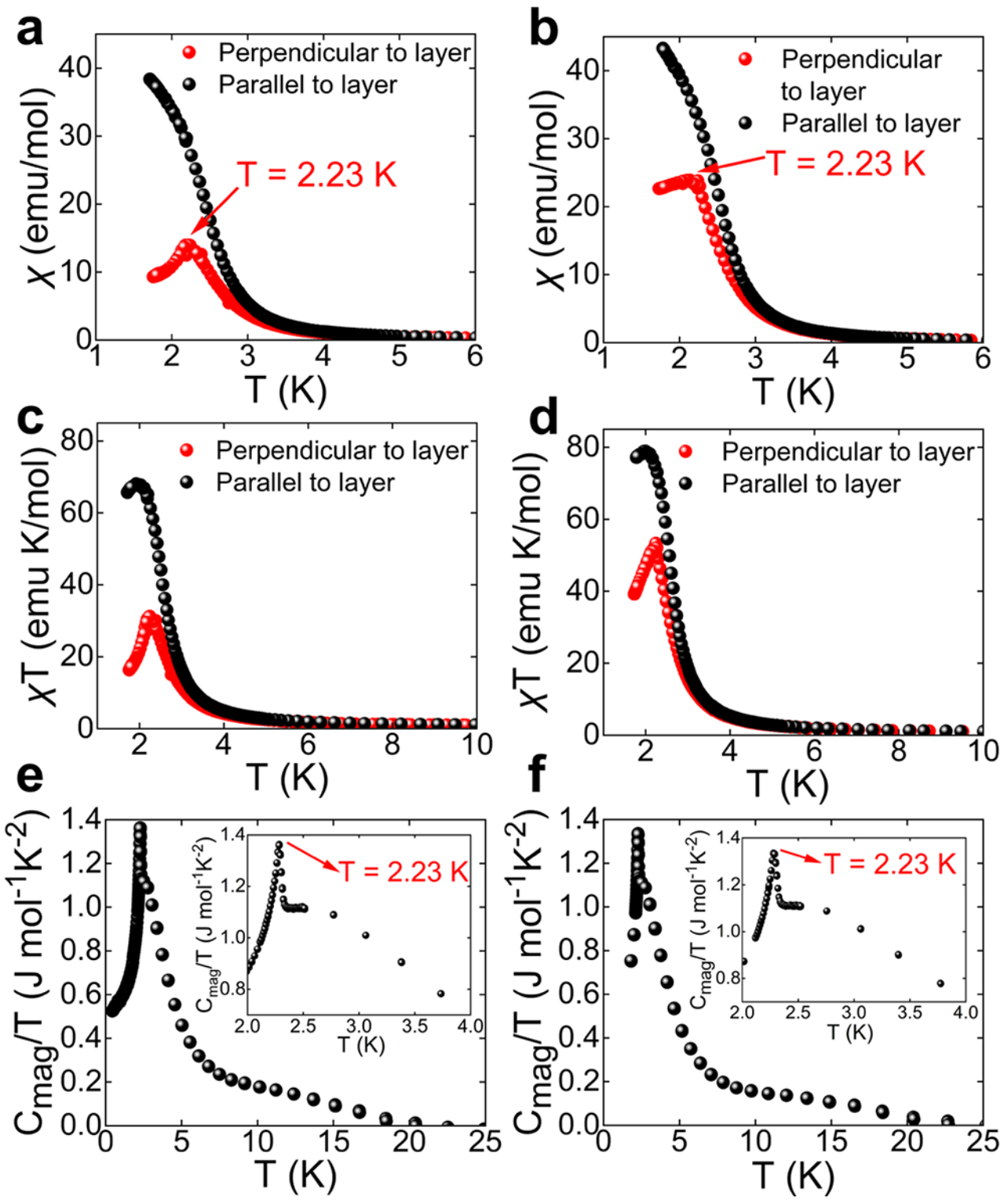

**Figure 4.** Plots of the magnetic susceptibility and magnetic specific heat capacity data for (a), (c),(e) chiral $(C_4H_9FN)_2CuCl_4$ and (b), (d),(f) racemic $(C_4H_9FN)_2CuCl_4$ single crystals.

The measured magnetic susceptibility data vs temperature (at an applied field of 100 Oe) shown in Figure 4a-4d show that both chiral and racemic $(C_4H_9FN)_2CuCl_4$ single crystals behave as A-type antiferromagnets,[63] where ferromagnetic behavior dominates when measured along the inorganic Cu−Cl layer propagation direction (*ab* plane for (R)-/(S)-$(C_4H_9FN)_2CuCl_4$, *bc* plane for racemic $(C_4H_9FN)_2CuCl_4$, see Figure 1), though we do not observe a ferromagnetic phase transition. As field-cooled measurements are plotted, a downturn in the susceptibility below the critical temperature would not be present for ferromagnets, supporting our hypothesis that the chiral and racemic variants are all A-type antiferromagnets. Our findings are consistent with the magnetic behavior of previously reported achiral Cu(II) 2D Ruddlesden−Popper phase (*i.e.*, Cu(II) ions in adjacent layers are staggered) perovskites.[64−66] A wide variety of magnetic behaviors have been reported in organic−inorganic hybrid 2D Cu(II) perovskites, where the magnetic behavior depends on the relative strengths of inter- and intralayer coupling as well as whether the Cu(II) ions in adjacent layers are staggered or eclipsed; the intralayer ferromagnetic coupling decreases as the length of the long Cu-X contact increases,[67,68] and the interlayer antiferromagnetic coupling decreases as the distance between Cu-X layers increases.[69] When measured perpendicular to the Cu−Cl layer propagation direction, an antiferromagnetic phase transition at a Néel temperature $T_N$ = 2.23 K is observed for both chiral and racemic $(C_4H_9FN)_2CuCl_4$. This transition is also observed in the heat capacity data (Figures 4e−f and S9), which supports the existence of a magnetic phase transition for chiral and racemic $(C_4H_9FN)_2CuCl_4$ at 2.23 K. To determine the magnetic entropy, we fit the heat capacity data from 20 to 300 K using the double Debye model (Figure S9) to account for the distribution of phonon energies in both the organic and inorganic moieties.[70,71] The fitting yields two Debye temperatures: 335.4 and 120.3 K (for chiral $(C_4H_9FN)_2CuCl_4$), and 383.2 and 130.2 K (for racemic $(C_4H_9FN)_2CuCl_4$), respectively. After subtracting the extrapolated double Debye model fit from the total heat capacity, we extract a magnetic entropy of ∼5.74 J·mol$^{-1}$·K$^{-1}$ and ∼4.94 J·mol$^{-1}$·K$^{-1}$, respectively, for chiral and racemic $(C_4H_9FN)_2CuCl_4$, both of which are close to the expected value ($R$ ln(2) = 5.76 J·mol$^{-1}$·K$^{-1}$) for magnetic order in an S = 1/2 material. When the magnetization is measured parallel to the Cu−Cl layer propagation direction, no antiferromagnetic phase transition is observed, with only a small downturn in the $\chi T$ plot below $T_N$ observed for the both the chiral and racemic variants. The presence of stronger ferromagnetic interactions in the single-crystal susceptibility measurements is consistent with the greater Weiss temperature of the racemic variant (Table 2).

Field-dependent magnetization measurements at 1.8 K are shown in Figure 5. Based on the measured data, as expected, both chiral and racemic $(C_4H_9FN)_2CuCl_4$ exhibit para- or ferromagnetic behavior, consistent with our hypothesis that these materials are A-type (easy-axis) antiferromagnets, as hysteresis is present. However, unlike the racemic sample, there is a hesitation in the change of magnetic moment around the zero magnetic field region for chiral $(C_4H_9FN)_2CuCl_4$ (Figure 5a). This "hesitation" in the change of magnetic moment is likely a spin-flip transition, where once a critical field is reached, the spins rapidly rearrange to align with one another, and the moment then quickly saturates.[45,72,73] A similar spin-flip transition has has been observed in $EA_2CuCl_4$ (EA = ethylammonium)[45] We note that this is distinct from a spin-flop transition, in which the transition is second order, where the moment slowly changes with increasing field after the sudden change in magnetization.[45,66,68] Spin-flop transitions are much more common than spin-flip transitions in 2D hybrid organic−inorganic magnetic metal halide materials because of the antiferromagnetic coupling, and examples of materials that exhibit spin-flop transitions are $EA_2CuBr_4$, $(PEA)_2MnCl_4$, and $(PEA)_2CuCl_4$ (where PEA = 2-phenethylammonium);[74−77] $EA_2CuCl_4$ exhibits both a spin-flip and spin-flop transition, depending on temperature and field.[45] In addition, spin-flip transitions are indicative of greater magnetocrystalline anisotropy than spin-flop transitions,[72,78] indicating that (R)- and (S)-$(C_4H_9FN)_2CuCl_4$ exhibit larger magneto-crystalline anisotropy than most other reported hybrid magnetic perovskites.

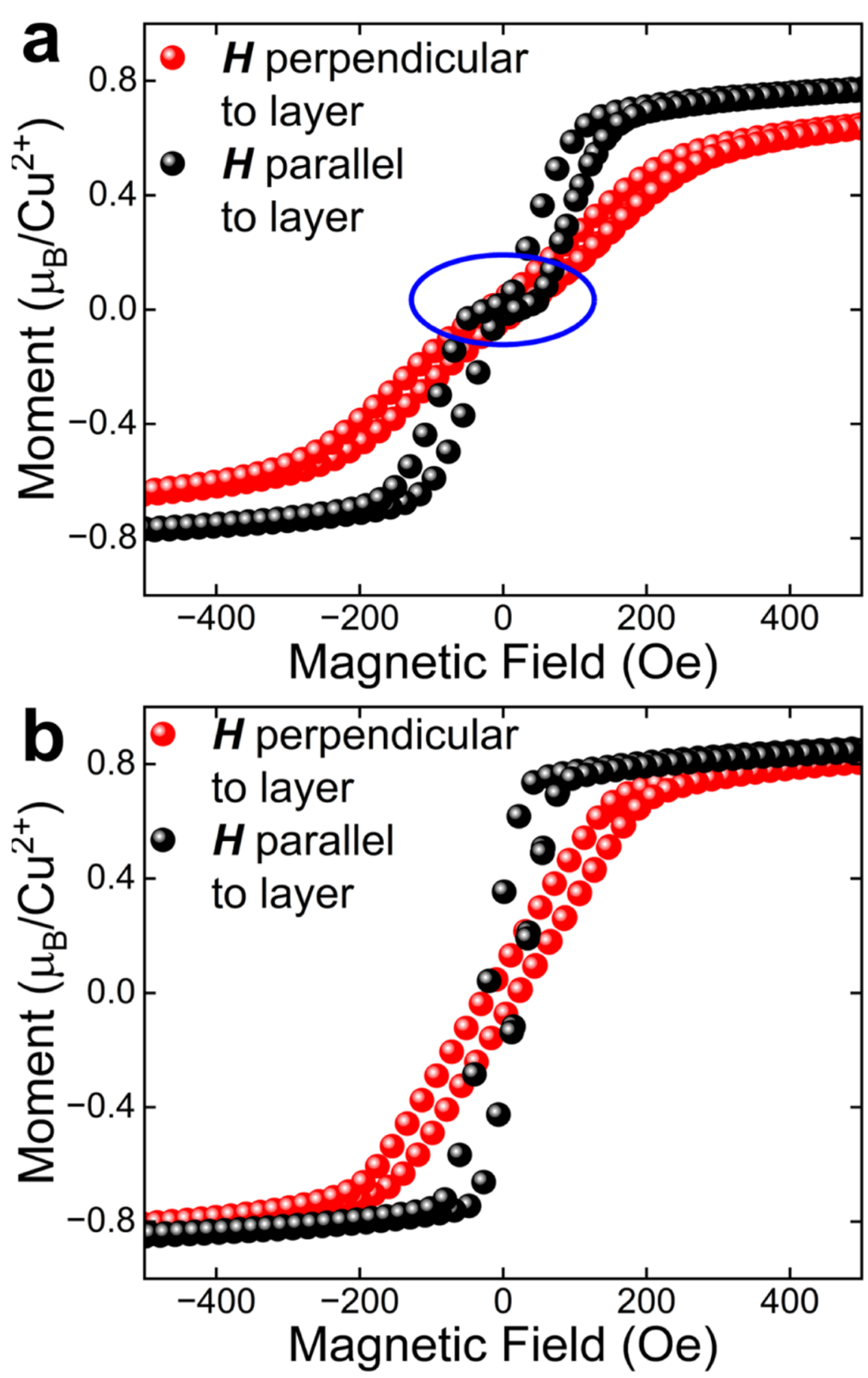

**Figure 5.** Plots of the magnetic moment vs magnetic field H (at $T$ = 1.8 K) for (a) chiral and (b) racemic $(C_4H_9FN)_2CuCl_4$ single crystals. The blue circle in panel (a) is likely a spin-flip transition.

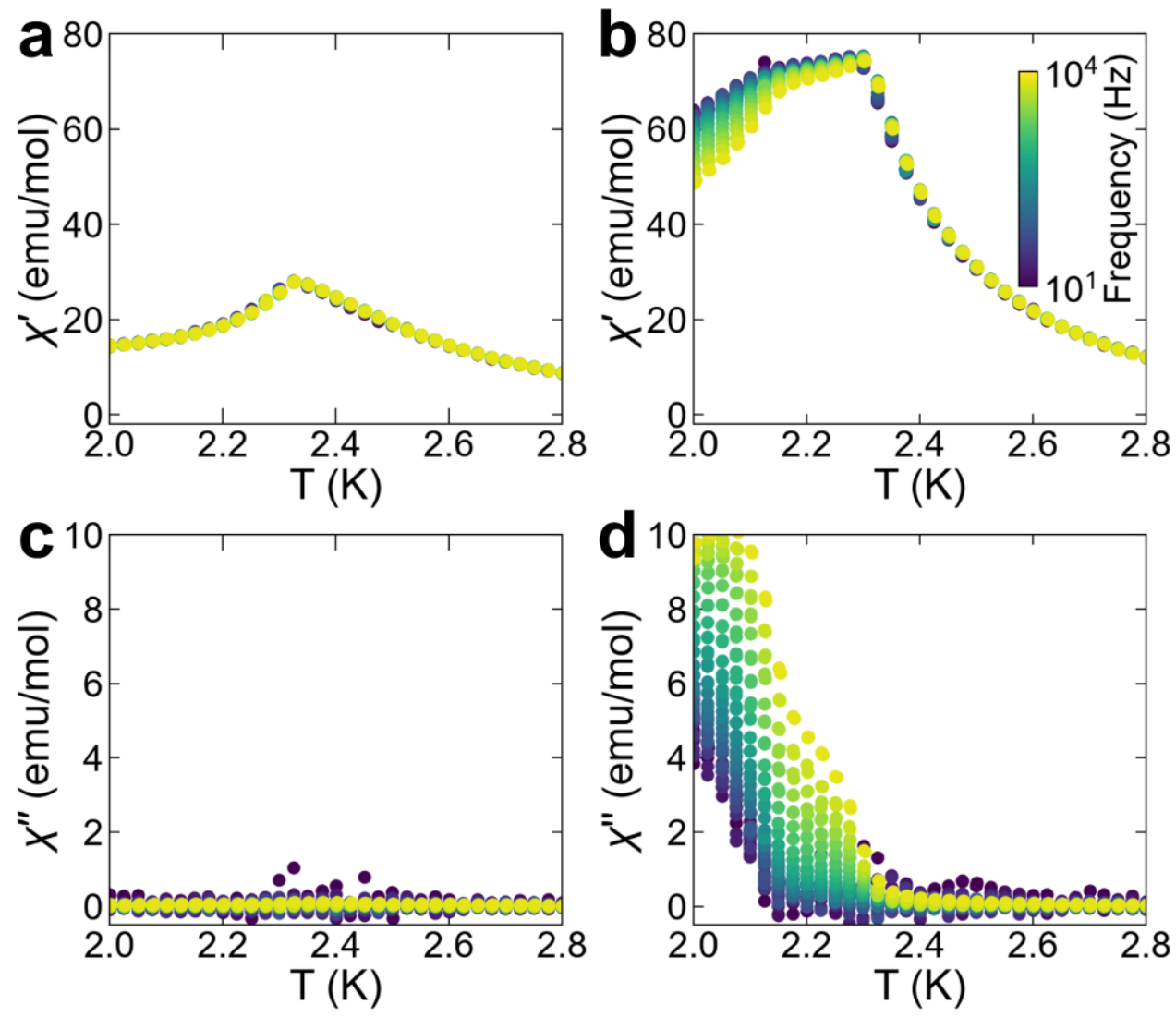

**Figure 6.** In-phase ($\chi'$) component of AC susceptibility for (a) (R)-$(C_4H_9FN)_2CuCl_4$ and (b) racemic $(C_4H_9FN)_2CuCl_4$. Out-of-phase ($\chi''$) component of AC susceptibility for (c) (R)-$(C_4H_9FN)_2CuCl_4$ and (d) racemic $(C_4H_9FN)_2CuCl_4$.

In addition to measuring the magnetic moment, we carried out frequency-dependent AC magnetization measurements for the chiral and racemic samples (Figures 6 and S10). Figure 6a and 6b present the temperature-dependent in-phase component ($\chi'$) of the AC magnetic susceptibility data for (R)- and racemic $(C_4H_9FN)_2CuCl_4$ powder, respectively, measured under a 1.5 Oe AC magnetic field. We once more observed the Néel temperature transition at 2.23 K in both the chiral and racemic materials. For chiral $(C_4H_9FN)_2CuCl_4$, the measured AC susceptibility data show no frequency dependence. However, for the racemic material, the moment decreases with increasing frequency below the Néel temperature (Figure 6b). Despite the frequency dependence of the susceptibility, the racemic variant does not form a spin glass because the Néel temperature is independent of frequency.[79,80] We hypothesize that the difference in frequency dependence between the chiral and racemic materials originates from the much higher strength of the antiferromagnetic interaction in the chiral material compared to the racemic, which can be observed in both the temperature- and field-dependent magnetic data (Figures 4 and 5), especially by the presence of the spin-flip transition in the chiral variant. At low field, antiferromagnetic behavior dominates in the chiral material regardless of the direction of the applied field, so the frequency-dependent hysteresis that would be expected for a ferromagnet is not observed. In contrast, the racemic variant has much weaker antiferromagnetic interactions despite having an identical Néel temperature, so hysteresis can be seen in the frequency dependence of the AC susceptibility, indicative of the presence of domain walls that are present in ferro- and ferrimagnets.[81] The behavior of the out-of-phase component of the AC susceptibility ($\chi''$) is also consistent with this explanation—it is approximately 0 for (R)- and (S)-$(C_4H_9FN)_2CuCl_4$ (Figures 6c and S10), indicative of antiferromagnetism, but shows a dramatic frequency dependence below ~2.3 K in racemic $(C_4H_9FN)_2CuCl_4$ (Figure 6d), indicative of domain-wall ferro- or ferrimagnetism. To conclusively demonstrate the presence of long-range magnetic order, neutron diffraction measurements must be performed.

Figure S11 presents the resistivity data for $(C_4H_9FN)_2CuCl_4$ samples. The crystal shows a resistivity of $\sim 10^4$ Ω·cm (measured along the Cu−Cl layer propagation direction), while, when measured perpendicular to the layer propagation direction, a resistivity of $\sim 10^7$ Ω·cm is obtained. The higher resistivity is consistent with the presence of organic cations between inorganic layers.

Magnetoelectric data are shown in Figure 7. Importantly, the chiral variant shows a magnetoelectric signal, demonstrating the existence of chiral magnetic order in the material. The chiral crystal structure (space group $P2_12_12_1$, no. 19) breaks spatial inversion symmetry and permits higher-order coupling between the lattice and magnetization. Specifically, when an in-plane magnetic field containing both $a$- and $b$-axis components is applied, the $a$-axis field, together with the structural chirality, effectively induces a toroidal moment along the $a$-axis. This induced toroidal moment, in combination with the magnetic field component along $b$, further generates an electric polarization along the $c$-axis, i.e., a second-order magnetoelectric effect.[5] Notably, this mechanism does not require long-range chiral magnetic order from a symmetry perspective (short-range chiral correlations are sufficient), but lower temperatures are beneficial as they suppress thermal fluctuations and enhance the magnetization response. As shown in Figure 7, the chiral variant shows a clear magnetoelectric signal at 2 K, with the current reversing its sign upon reversing the magnetic field, consistent with the presence of a toroidal moment and the chirality-induced second-order magnetoelectric mechanism described above. In contrast, the racemic sample shows no detectable magnetoelectric response when an in-plane magnetic field (here, containing both *b*- and *c*-axis components) is applied, indicating the absence of a field-induced toroidal moment, as expected for an achiral material.

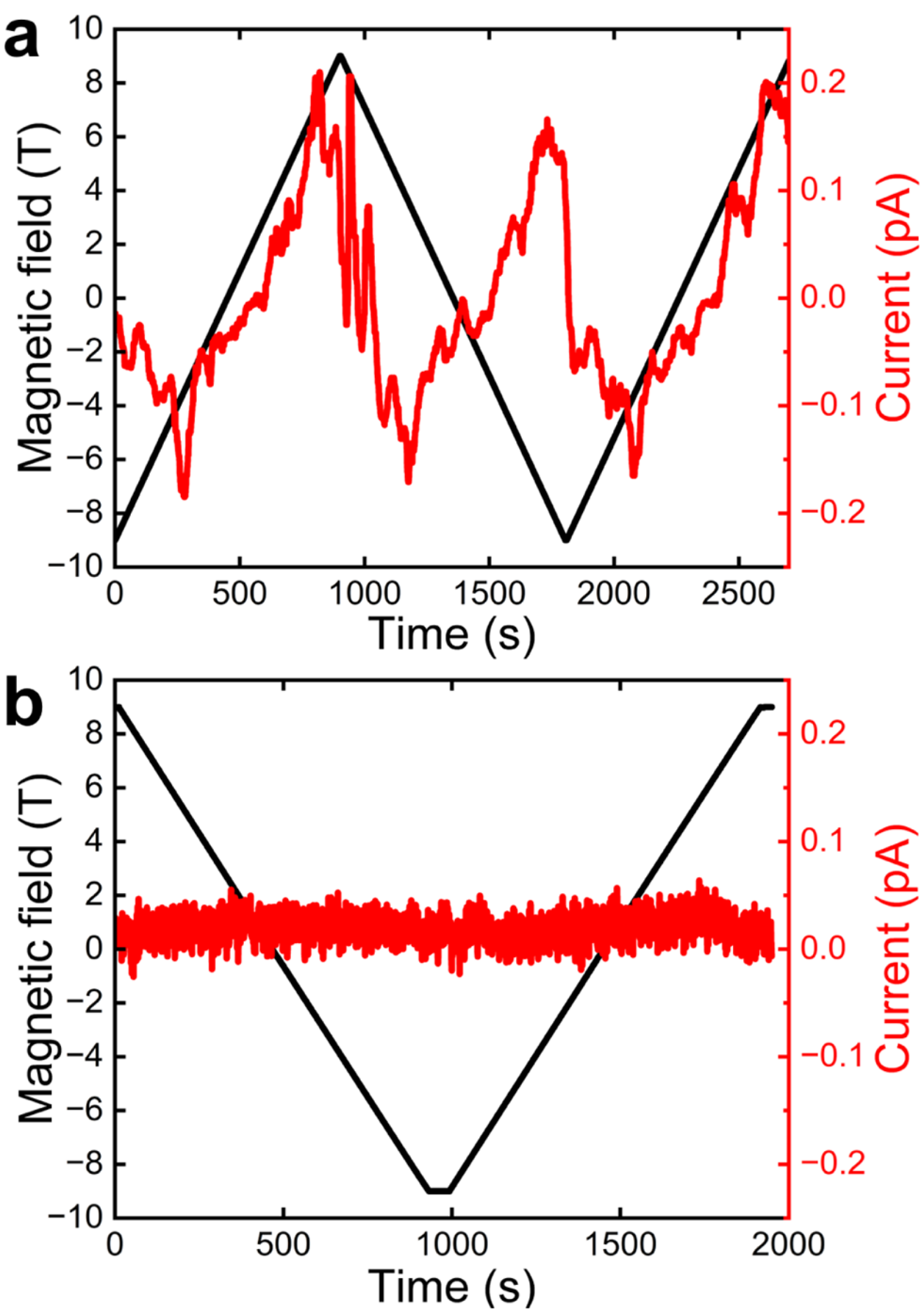

**Figure 7.** Magnetoelectric data collected at 2 K on (a) chiral and (b) racemic $(C_4H_9FN)_2CuCl_4$ crystals. The magnetic field is applied along the in-plane direction, and the out-of-plane magnetoelectric current is collected.

## 3. CONCLUSION

To summarize, we demonstrate that in the chiral organic−inorganic hybrid perovskite $(C_4H_9FN)_2CuCl_4$, the chiral organic cations induce the formation of chiral magnetic order in the nearly centrosymmetric magnetic sublattice. For both chiral and racemic $(C_4H_9FN)_2CuCl_4$, a magnetic phase transition at $T_N$ = 2.23 K is observed when measured perpendicular to the Cu−Cl layer propagation direction, and specific heat capacity measurements reveal a magnetic entropy consistent with S = 1/2 magnetic order for both chiral and racemic $(C_4H_9FN)_2CuCl_4$. The strength of the interlayer antiferromagnetic interactions is stronger in the chiral variants than in the racemic variant. Finally, the presence of the magnetoelectric effect in chiral $(C_4H_9FN)_2CuCl_4$ supports the presence of magnetic chirality in (R)- and (S)-$(C_4H_9FN)_2CuCl_4$, while the lack of a magnetoelectric effect in racemic $(C_4H_9FN)_2CuCl_4$ emphasizes that the structural chirality of (R)- and (S)-$(C_4H_9FN)_2CuCl_4$ is responsible for the toroidal magnetic signal. These findings demonstrate the importance of the chiral organic cations in the magnetic behavior of organic−inorganic hybrid materials, warranting additional study to determine the magnetic crystal structures of these and other chiral magnetic hybrid materials and their potential application in magnetic information processing and spin−orbit torque memory devices. Furthermore, the combination of magnetic chirality and large magnetocrystalline anisotropy suggests that the properties of (R)- and (S)-$(C_4H_9FN)_2CuCl_4$ may provide insight into the origin of the CISS effect and its relation to asymmetric Dzyaloshinskii−Moriya interactions that are responsible for chiral magnetism.[82]

## ■ ASSOCIATED CONTENT

### Supporting Information

The Supporting Information is available free of charge at https://pubs.acs.org/doi/10.1021/jacs.6c11636.

> Experimental section, additional depictions of crystal structures, powder X-ray diffraction (PXRD) patterns, linear dichroism and linear birefringence correction to circular dichroism spectra, plot of the anisotropy factor ($g_{CD}$), additional susceptibility data on powder samples, additional heat capacity data, additional AC susceptibility data, electronic measurement results, output of PLATON ADDSYM (PDF)

### Accession Codes

Deposition Numbers 2543946−2543948 contain the supplementary crystallographic data for this paper. These data can be obtained free of charge via the joint Cambridge Crystallographic Data Centre (CCDC) and Fachinformationszentrum Karlsruhe Access Structures service.

## ■ AUTHOR INFORMATION

### Corresponding Author

**Daniel B. Straus** − *Department of Chemistry, Tulane University, New Orleans, Louisiana 70118, United States;* orcid.org/0000-0003-2977-5590; Email: dstraus@tulane.edu

### Authors

**Zheng Zhang** − *Department of Chemistry, Tulane University, New Orleans, Louisiana 70118, United States*

**Mingyu Xu** − *Department of Chemistry, Michigan State University, East Lansing, Michigan 48824, United States*

**Jose L. Gonzalez Jimenez** − *Department of Chemistry, Michigan State University, East Lansing, Michigan 48824, United States*

**Stephen Zhang** − *Department of Chemistry, Princeton University, Princeton, New Jersey 08540, United States*

**Weiwei Xie** − *Department of Chemistry, Michigan State University, East Lansing, Michigan 48824, United States*

**Xianghan Xu** − *School of Physics and Astronomy, University of Minnesota, Twin Cities, Minnesota 55455, United States;* orcid.org/0000-0001-6854-300X

Complete contact information is available at: https://pubs.acs.org/10.1021/jacs.6c11636

## Notes

The authors declare no competing financial interest.

## ■ ACKNOWLEDGMENTS

Primary support for this work is provided by the Louisiana Board of Regents under contract LEQSF(2024-27)-RD-A-28. The authors thank Dr. Xiaodong Zhang for his assistance with single-crystal XRD measurements, Dr. Alex McSkimming and Addie Fraker for assistance with UV−vis absorbance data collection, Dr. Huibo Cao and Dr. Yiqing Hao for helpful discussions, and Sienna Alonso for assistance with thin-film preparation. Z. Z. and D. S. acknowledge additional financial support from Tulane University. S. Z. is supported by the U.S. Department of Energy under Grant No. DE-FG02-98ER45706. J. G., M. X., and W. X. are supported by the U.S. Department of Energy under Grant No. DE-SC0024943. X. X. acknowledges support from the U.S. Department of Energy through the University of Minnesota Center for Quantum Materials under Grant No. DE-SC-0016371.

## ■ REFERENCES

# Supporting Information

## Chirality Transfer to the Magnetic Sublattice in the Hybrid Perovskite (R)-/(S)-3-Fluoropyrrolidinium Copper(II) Chloride

Zheng Zhang[a], Mingyu Xu[b], Jose L. Gonzalez Jimenez[b], Stephen Zhang[c], Weiwei Xie[b], Xianghan Xu[d], and Daniel B. Straus[a*]

[a]Department of Chemistry, Tulane University, New Orleans, LA, USA 70118
[b]Department of Chemistry, Michigan State University, East Lansing, MI, USA 48824
[c]Department of Chemistry, Princeton University, Princeton, NJ, USA 08540
[d]School of Physics and Astronomy, University of Minnesota, Twin Cities, MN, USA 55455
(*Author to whom correspondence should be addressed: dstraus@tulane.edu)

## 1. Experimental Section

### Materials

Copper (II) chloride (anhydrous, 99% extra pure) was purchased from Acros Organics. (R)-3-fluoropyrrolidine hydrochloride and (S)-3-fluoropyrrolidine hydrochloride were purchased from Ossila and Ambeed, respectively. Hydrochloric acid (36.5-38%) was purchased from VWR International. Dimethylformamide (DMF, ≥99.8%, purity grade: spectranalyzed) was purchased from Fisher Scientific.

### Crystal Growth

To synthesize (R)- or (S)-$(C_4H_9FN)_2CuCl_4$ crystals, 2 mmol (R)-3-fluoropyrrolidine hydrochloride or 2 mmol (S)-3-fluoropyrrolidine hydrochloride was first dissolved in 3 mL HCl acid. Then, 2 mmol $CuCl_2$ was added to the solution. The solution was then vigorously stirred until a clear solution was obtained. For the synthesis of racemic $(C_4H_9FN)_2CuCl_4$ crystals, 1 mmol (R)-3-fluoropyrrolidine hydrochloride and 1 mmol (S)-3-fluoropyrrolidine hydrochloride were added to 3 mL HCl acid. 2 mmol $CuCl_2$ was then added, and a clear solution was obtained after stirring. Crystal growth experiments were performed by the evaporation of solvent from the prepared solutions in open air. Typically, mm-to-cm crystals scale are obtained after one week.

### X-Ray Diffraction (XRD) Measurements

Single crystal X-ray diffraction data were collected at room temperature using a Bruker D8 Quest diffractometer equipped with a Photon III 7 detector using Mo Kα radiation or a Bruker D8 Venture diffractometer equipped with a Photon III 14 detector using Cu Kα X-ray irradiation. The crystal structures were solved using ShelXT[1] and refined using ShelXL[2] within the Olex2 GUI.[3]

For powder X-ray diffraction measurements, (R)-, (S)-, or racemic $(C_4H_9FN)_2CuCl_4$ crystals were ground into a powder using a mortar and pestle, which was measured using a Malvern Panalytical Aeris diffractometer equipped with a PIXcel3D detector using Cu Kα radiation. To check stability against ambient air exposure, $(C_4H_9FN)_2CuCl_4$ powders were stored under ambient conditions in the dark, and periodic PXRD measurements were taken to monitor the PXRD pattern change.

### Thin Film Preparation

(R)-, (S)-, and racemic $(C_4H_9FN)_2CuCl_4$ thin films were spin-cast onto 25x25x1mm glass substrates. Substrates were cleaned by successive sonication in the following solvents or solutions: 5% v/v aqueous solution of Hellmanex III detergent in deionized water; deionized water (3x); 1 M hydrochloric acid; deionized water (3x); and ethanol (3x). The substrates were then treated in a UV-Ozone cleaner located in a nitrogen glove box for 30 min.

To prepare the films, 30 mg of (R)-, (S)-, or racemic $(C_4H_9FN)_2CuCl_4$ crystals were dissolved in 0.4 mL methanol. The solutions were spin-cast onto the cleaned glass substrates at 1000 RPM for 45 s in a nitrogen glove box.

### UV-Vis Absorbance and Circular Dichroism (CD) Measurements

UV-Vis absorbance data were collected on thin films of (R)-, (S)-, and racemic $(C_4H_9FN)_2CuCl_4$ using an Agilent Cary 50 Bio UV-Vis spectrometer in transmission. CD spectra were collected using an OLIS DSM 1000 CD spectrometer in transmission on the same thin films used in absorbance measurements. The CD spectrometer was calibrated using *d*-camphorsulfonic acid as a standard.[4] To account for linear dichroism (LD) and linear birefringence (LB) effects, the CD spectrum for each sample was measured twice (Figure S6), once with the film facing towards the detector and once with the film facing away from the detector. Averaging the two spectra gives the CD spectrum corrected for LD and LB effects.[5–7]

## Electrical Characterization

For current-voltage (I-V) and space-charge-limited-current (SCLC) measurements, conductive silver paste (purchased from Ted Pella, Inc., PELCO conductive silver paint) was used to attach copper wires to the top and bottom of a (R)-$(C_4H_9FN)_2CuCl_4$ crystal (vertical-type device structure). The crystal was then placed in an in-house built sample measurement box, which is connected to a Keithley 2636A sourcemeter. The I-V and SCLC testing were performed at room temperature and in a dark environment. For photo-response measurement, crystal was exposed to 365 nm light illumination (wattage 5 watts, bulb voltage 5 volts). The light was turned on and off to measure the crystal's response to light exposure.

## Magnetic Characterization

A Quantum Design Dynacool PPMS equipped with the ACMS II magnetization accessory was used to measure the temperature- and field-dependent DC magnetization of single crystal and powder samples of (R)-, (S)-, and racemic $(C_4H_9FN)_2CuCl_4$ as well as AC magnetization on powder samples. The temperature-dependent magnetization was measured at an applied field of 100 Oe on single crystal samples and 1000 Oe on powder samples. The susceptibility $\chi$ is defined as $M/H$, where M is the magnetization and H the applied field.

## Heat Capacity Measurements

Heat capacity measurements between 1.8 and 300 K were carried out using the Quantum Design Dynacool PPMS Heat Capacity accessory. Heat capacity measurements below 1.8 K used a Quantum Design $^3$He dilution refrigerator insert for the PPMS.

## Magnetoelectricity Measurements

Chiral and racemic crystals were cleaved into plate-like samples with a thickness of 0.5(1) mm. Silver paste was applied to form top and bottom electrodes. Due to thermal decomposition of the material, the electrodes were cured at room temperature over an extended period. The electrode areas for both devices are 2.0(5) mm$^2$. The devices were then cooled to 2 K in a PPMS under a 9 T in-plane magnetic field applied along (110) for chiral crystals and (011) for racemic crystals, and an applied out-of-plane electric field of 2 kV/cm. Magnetoelectric current signals at 2 K were measured using a Keithley 617 electrometer while sweeping the in-plane magnetic field at a rate of 200 Oe/s.

## 2. Additional Depictions of Crystal Structures

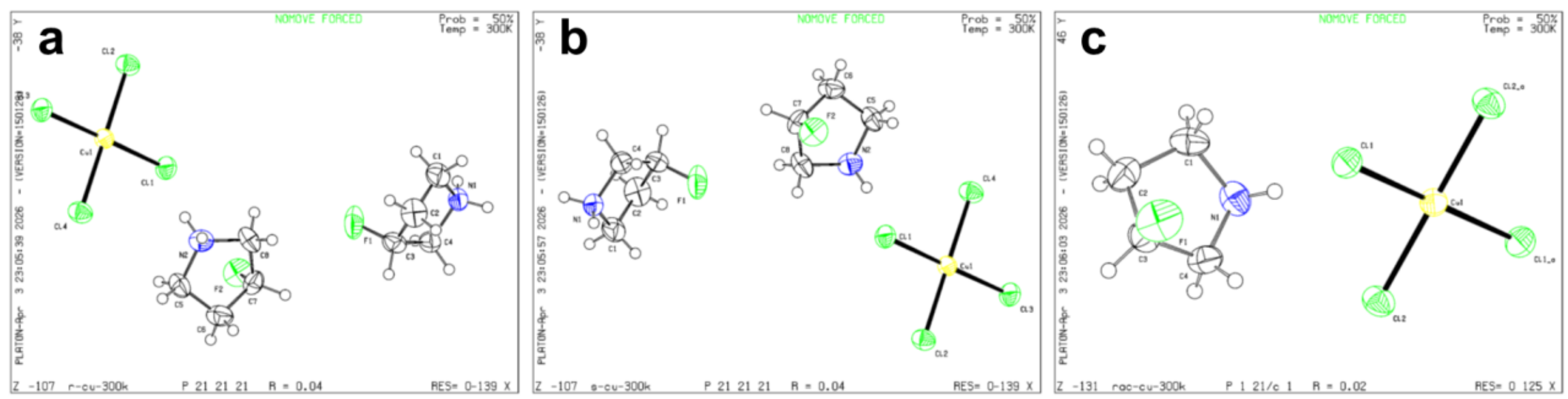

**Figure S1.** Asymmetric units of **(a)** (R)-, **(b)** (S)-, and **(c)** racemic $(C_4H_9FN)_2CuCl_4$ crystal structures, with atoms represented as 50% probability thermal ellipsoids.

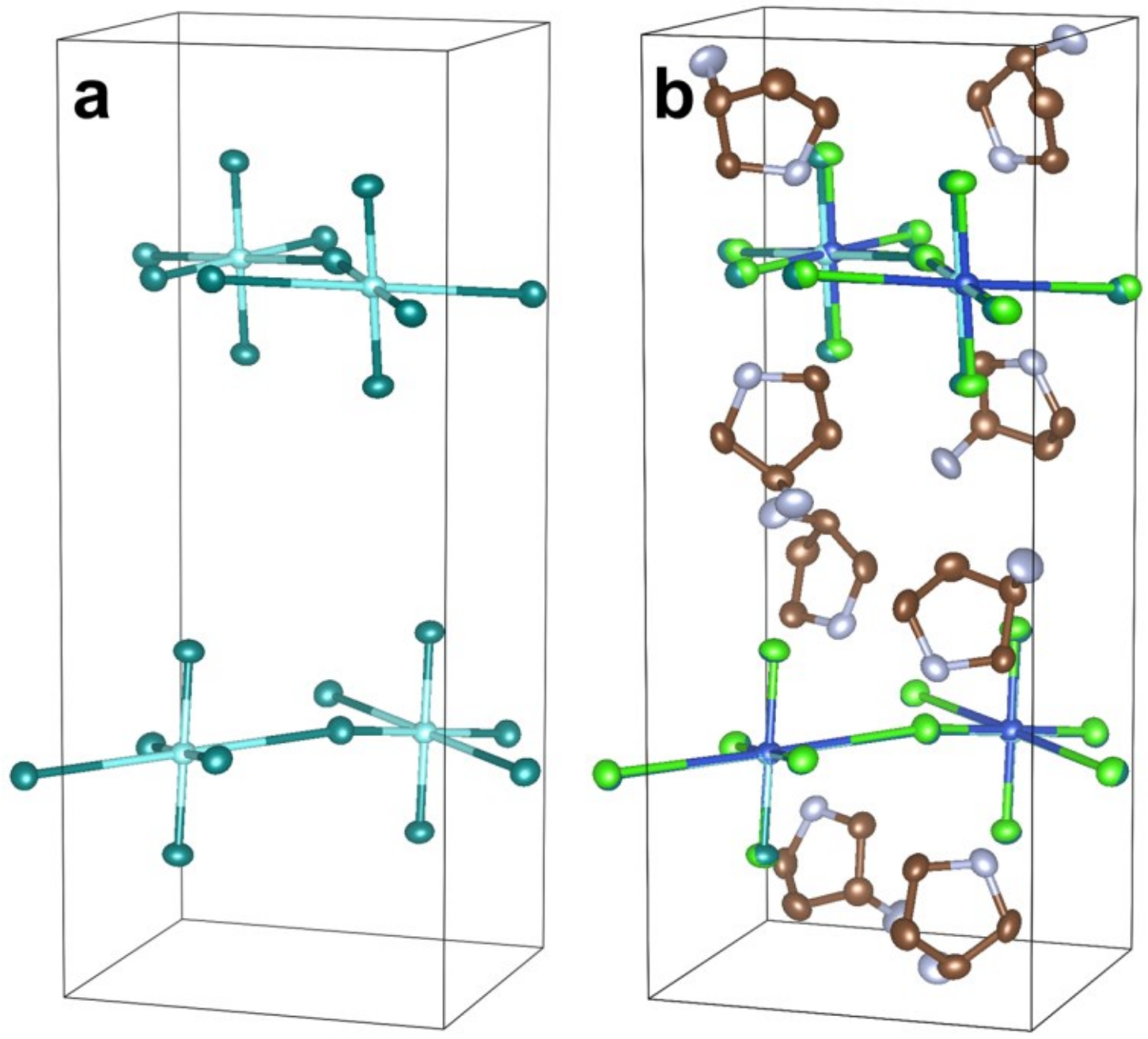

**Figure S2. (a)** Centrosymmetric inorganic framework of (R)-$(C_4H_9FN)_2CuCl_4$ ~~(R)-fluoropyrrolidinium copper(II)~~ output by ADDSYM, and **(b)** experimental (chiral) crystal structure rendered on top of centrosymmetric framework shown in (a). H atoms omitted in (b) for clarity.

## 3. Powder X-Ray Diffraction (PXRD) Pattern

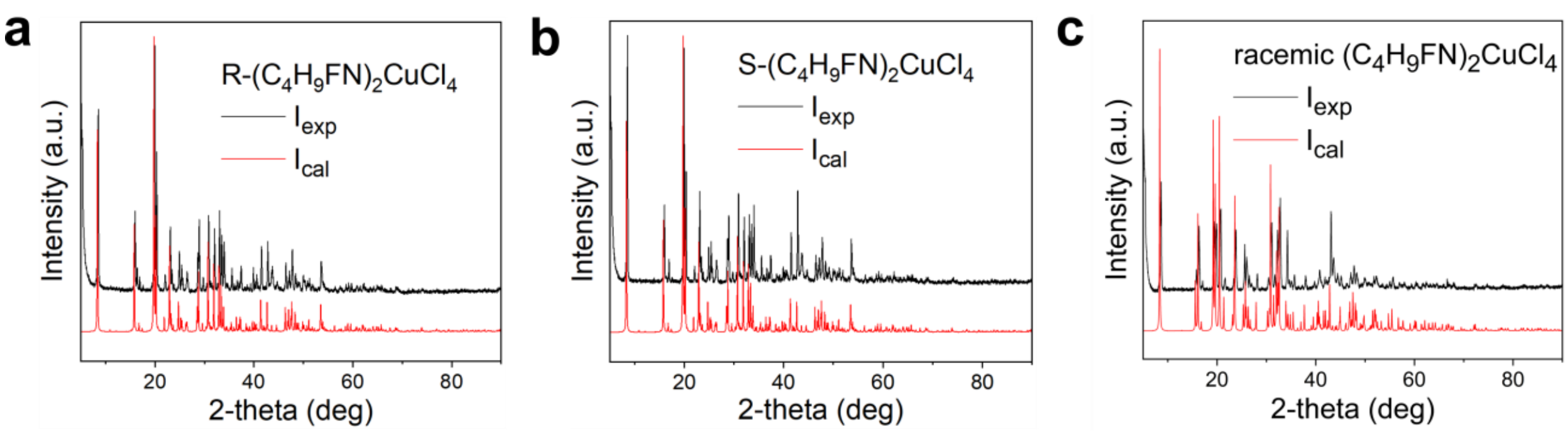

**Figure S3.** Experimental ($I_{exp}$) and calculated ($I_{cal}$) powder X-ray diffraction pattern of **(a)** (R)-($C_4H_9FN)_2CuCl_4$, **(b)** (S)-($C_4H_9FN)_2CuCl_4$, and **(c)** racemic ($C_4H_9FN)_2CuCl_4$ powders.

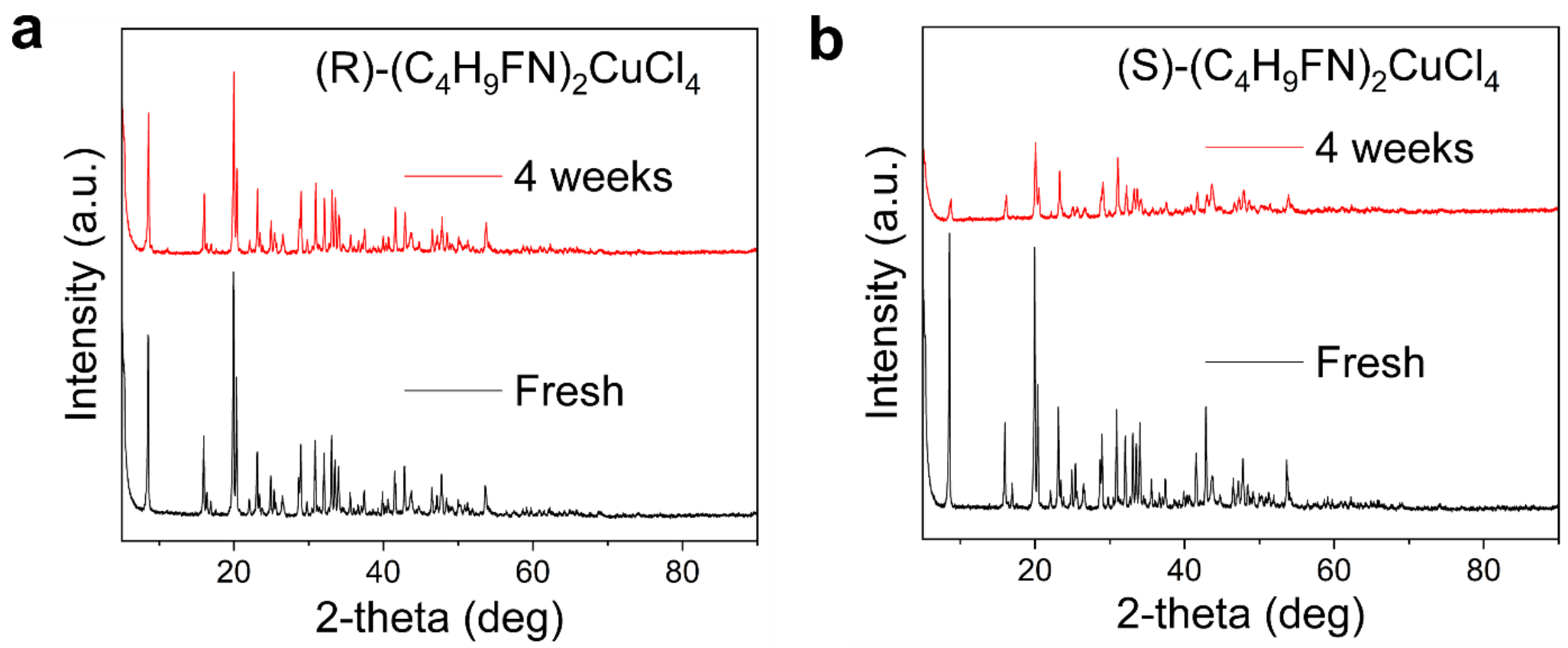

**Figure S4.** PXRD pattern measured for **(a)** (R)- and **(b)** (S)-($C_4H_9FN)_2CuCl_4$ to show their stability against ambient air exposure.

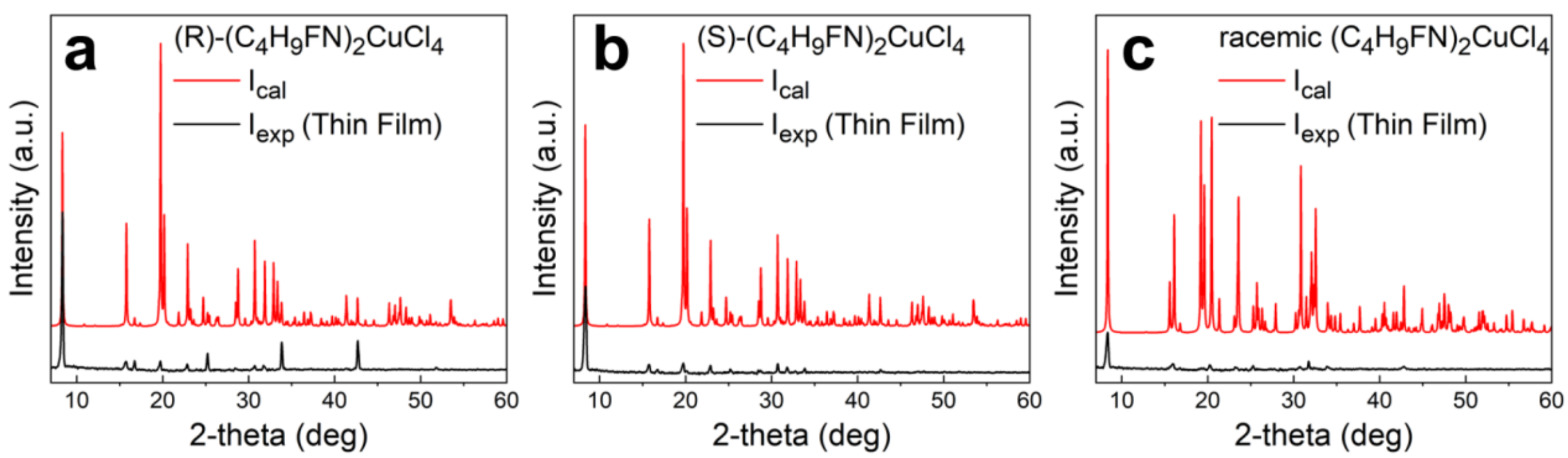

**Figure S5.** Experimental ($I_{exp}$) and calculated ($I_{cal}$) X-ray diffraction pattern of **(a)** (R)-($C_4H_9FN)_2CuCl_4$, **(b)** (S)-($C_4H_9FN)_2CuCl_4$, and **(c)** racemic ($C_4H_9FN)_2CuCl_4$ thin films.

## 4. Linear Dichroism and Linear Birefringence Correction to Circular Dichroism Spectra

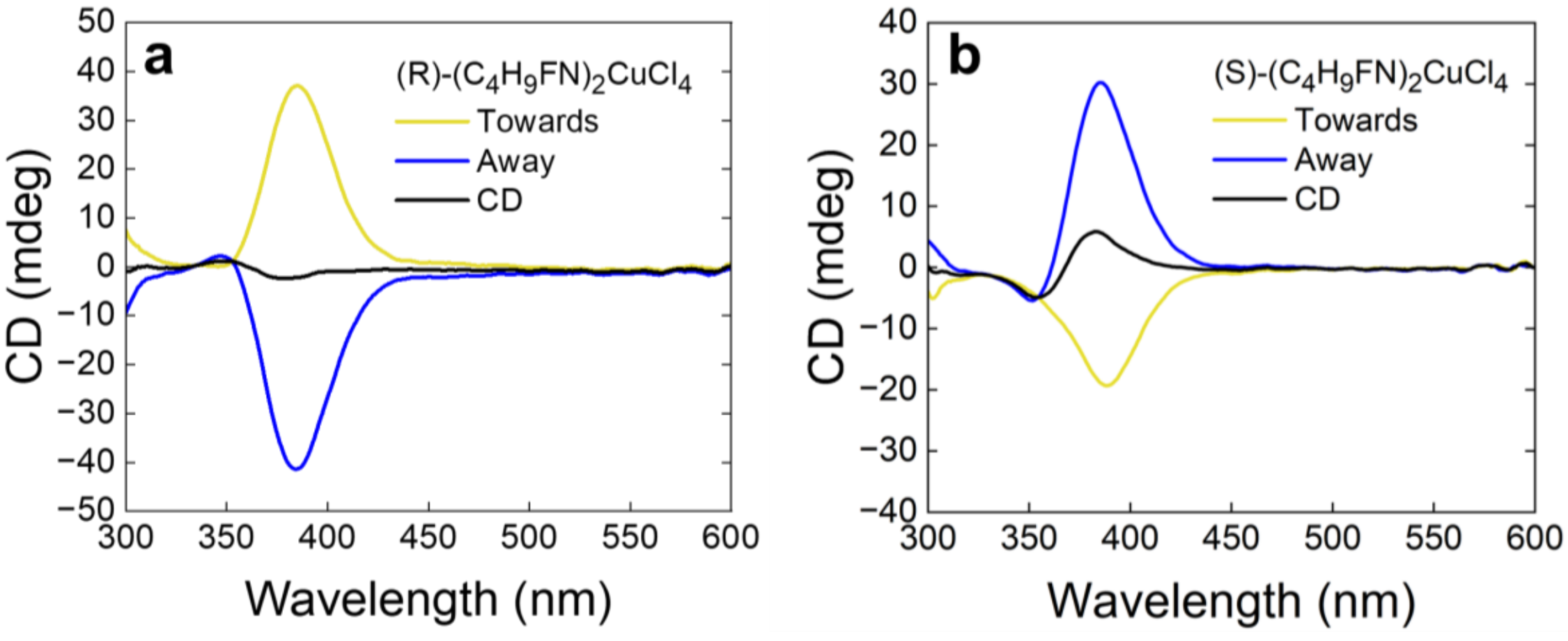

**Figure S6.** CD spectra of **(a)** (R)- and **(b)** (S)-$(C_4H_9FN)_2CuCl_4$ with (blue) film facing towards the detector, (yellow) away from the detector, and (black) true CD.

## 5. Plot of the Anisotropy Factor ($g_{CD}$)

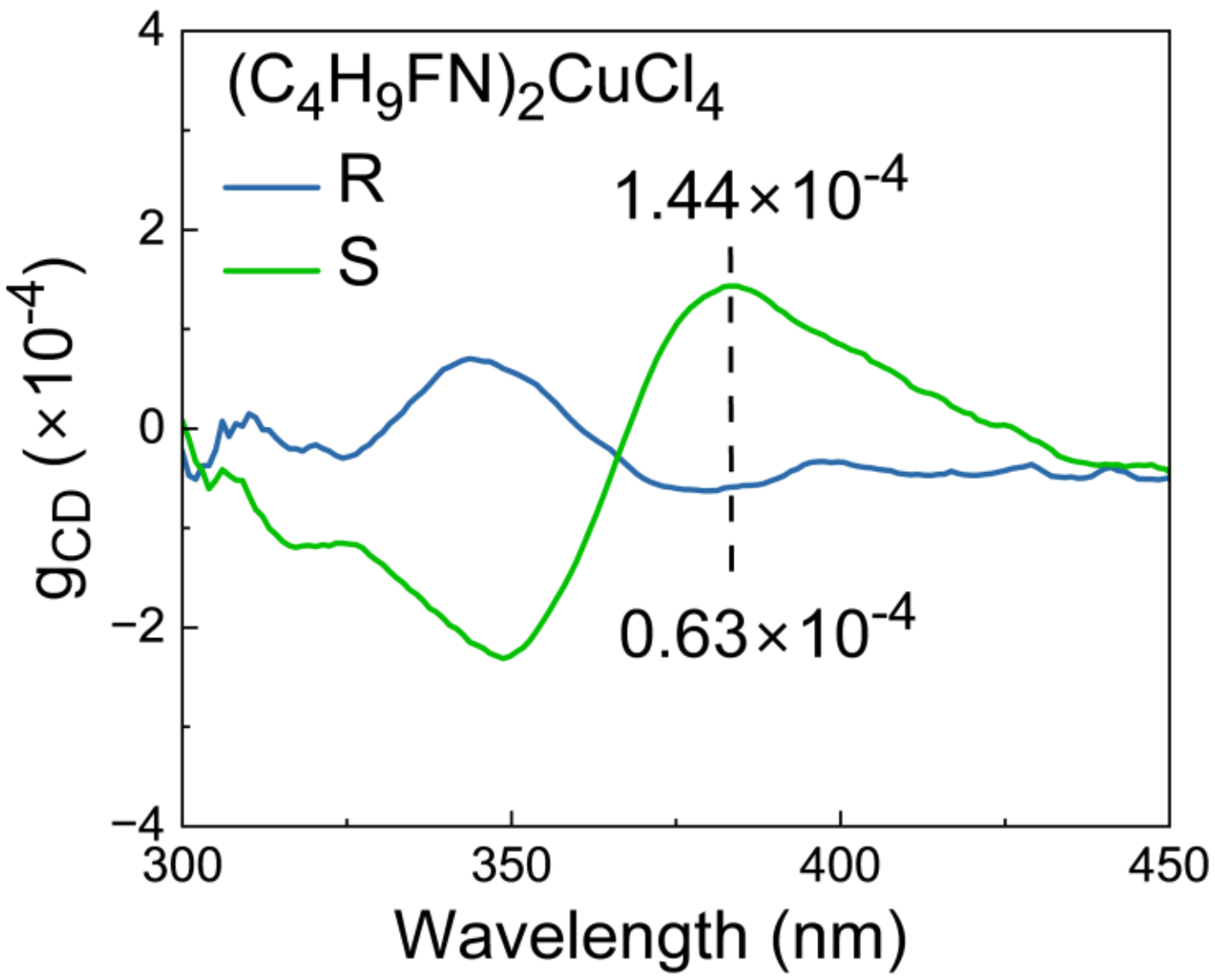

**Figure S7.** Plot of the anisotropy factor ($g_{CD}$) for (R)- and (S)-$(C_4H_9FN)_2CuCl_4$ thin films.

## 6. Additional Susceptibility Data on Powder Samples

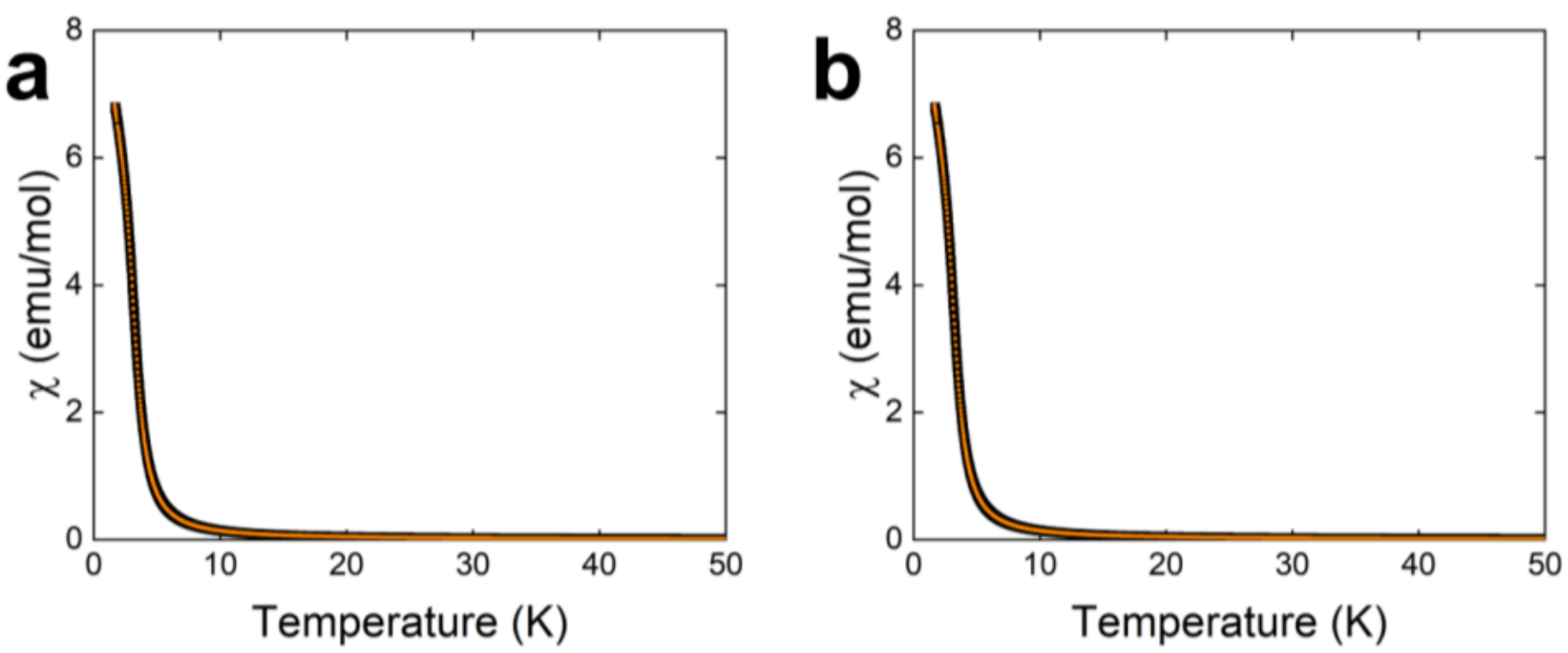

**Figure S8.** Plot of (black) zero-field cooled and (orange) field-cooled susceptibility data for **(a)** chiral $(C_4H_9FN)_2CuCl_4$ and **(b)** racemic $(C_4H_9FN)_2CuCl_4$ powder samples.

## 7. Heat Capacity Data

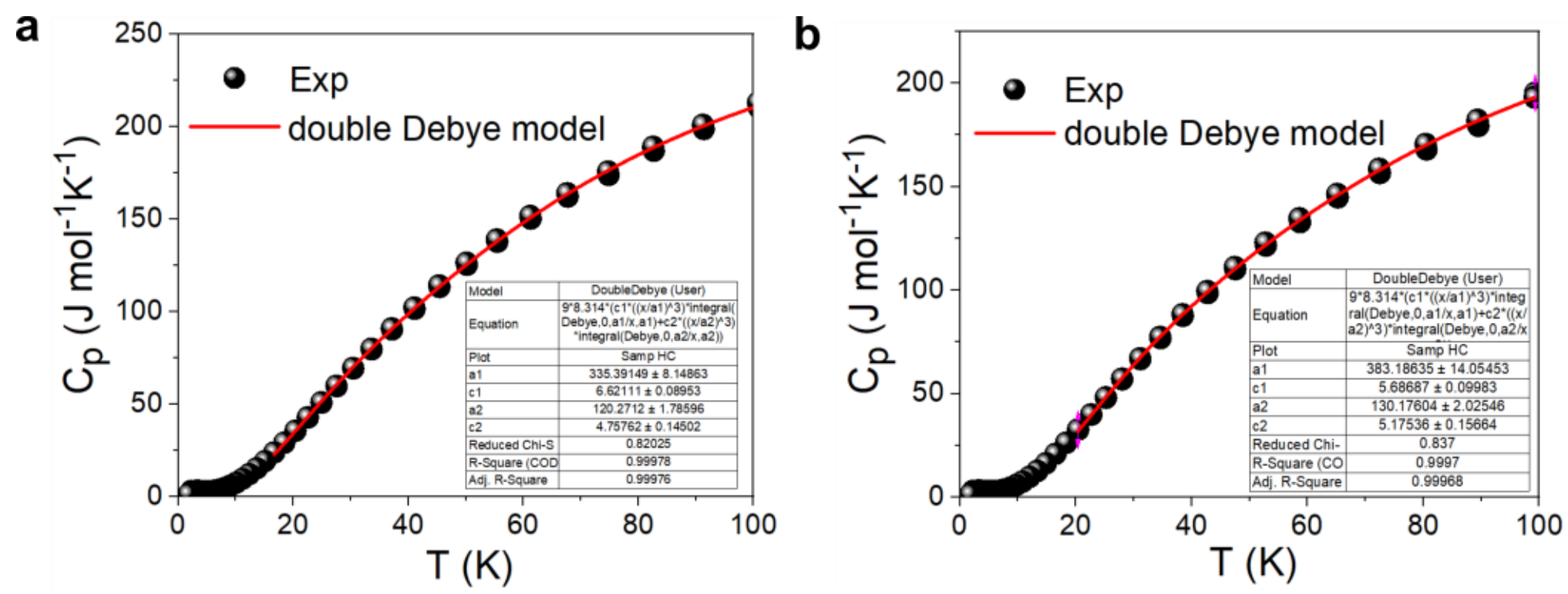

| Model | DoubleDebye (User) |
| --- | --- |
| Equation | 9*8.314*(c1*((x/a1)^3)*integral(Debye,0,a1/x,a1)+c2*((x/a2)^3)*integral(Debye,0,a2/x,a2)) |
| Plot | Samp HC |
| a1 | 335.39149 ± 8.14863 |
| c1 | 6.62111 ± 0.08953 |
| a2 | 120.2712 ± 1.78596 |
| c2 | 4.75762 ± 0.14502 |
| Reduced Chi-S | 0.82025 |
| R-Square (COD | 0.99978 |
| Adj. R-Square | 0.99976 |

| Model | DoubleDebye (User) |
| --- | --- |
| Equation | 9*8.314*(c1*((x/a1)^3)*integral(Debye,0,a1/x,a1)+c2*((x/a2)^3)*integral(Debye,0,a2/x |
| Plot | Samp HC |
| a1 | 383.18635 ± 14.05453 |
| c1 | 5.68687 ± 0.09983 |
| a2 | 130.17604 ± 2.02546 |
| c2 | 5.17536 ± 0.15664 |
| Reduced Chi- | 0.837 |
| R-Square (CO | 0.9997 |
| Adj. R-Square | 0.99968 |

**Figure S9.** Plot of the heat capacity data vs. T for **(a)** chiral $(C_4H_9FN)_2CuCl_4$ and **(b)** racemic $(C_4H_9FN)_2CuCl_4$ single crystals.

## 8. AC Susceptibility Data

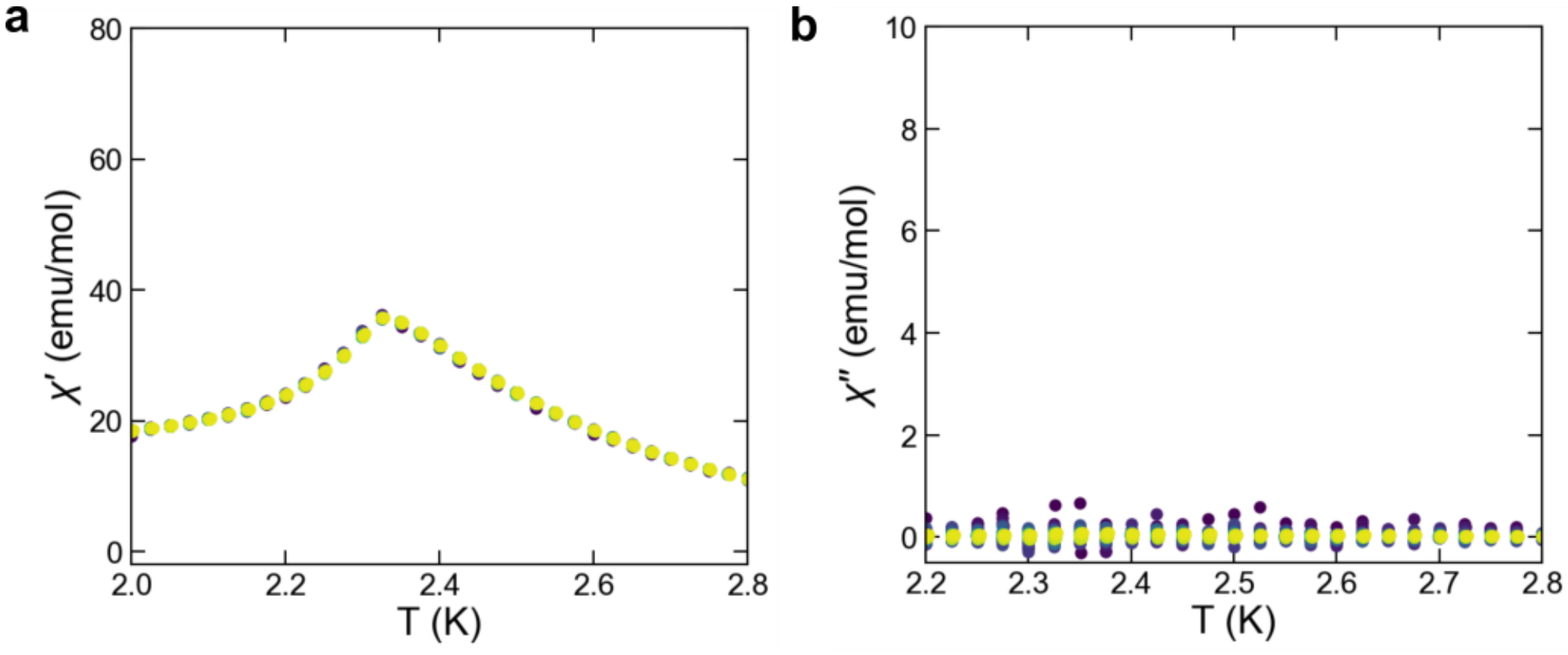

**Figure S10. (a)** In-phase ($\chi'$) component and **(b)** out-of-phase ($\chi''$) component of AC susceptibility for (S)-$(C_4H_9FN)_2CuCl_4$.

## 9. Electronic Measurement Results

Figure S11a shows the current-voltage (I-V) measurement data using as-grown (R)-($C_4H_9FN)_2CuCl_4$ crystals. Measured along the two-dimensional (2D) layer propagation direction (ab plane), a resistivity of ~$10^4$ $\Omega$·cm is obtained. Perpendicular to the 2D layer propagation direction (c axis), we obtained a resistivity of ~$10^7$ $\Omega$·cm. As expected, 2D metal halides show anisotropic electrical behaviors due to the insulating organic spacers that separate the metal-halide layers, resulting in a much greater resistivity when measured perpendicular to the layer direction. Similarly, the 2D Ruddlesden-Popper perovskite $(BA)_2CsPb_2Br_7$ (where BA is butylamine) has a resistivity of $2.22\times10^9$ $\Omega$·cm and $9.87\times10^{10}$ $\Omega$·cm along the ab plane and c axis, respectively.[8] In addition, using the space-charge-limited-current (SCLC) method,[9] we were able to estimate the density of trap states and carrier mobility in (R)-($C_4H_9FN)_2CuCl_4$ crystals from the trap-filled-limited (TFL) and Child regimes, respectively (Figure S11b). The density of trap states $n_t$ is determined using the formula below,[9]

$$n_t = \frac{2\varepsilon\varepsilon_0}{eL^2} V_{TFL}$$

where $\varepsilon$ (=13.02) is the dielectric constant, $\varepsilon_0$ is the vacuum permittivity, e is the electronic charge, L (=1 mm) is the crystal thickness, and $V_{TFL}$ is the onset voltage of the TFL regime. The density of trap states is determined to be $4.0\times10^{10}$ $cm^{-3}$ for (R)-($C_4H_9FN)_2CuCl_4$. This value is comparable to other metal halides (e.g., $2.77\times10^{11}$ $cm^{-3}$ for one-dimensional $CsCu_2I_3$, $3.33\times10^{10}$ $cm^{-3}$ for 2D $Rb_4Ag_2BiBr_9$.[10,11] For the charge carrier mobility, it is determined according to the Mott-Gurney law,[12]

$$\mu = \frac{8J_D L^3}{9\varepsilon\varepsilon_0 V^2}$$

where $\mu$ is the carrier mobility, $J_D$ is the current density at bias voltage V and L is again the crystal thickness. The carrier mobility is calculated to be 0.45 $cm^2$/V-s for (R)-($C_4H_9FN)_2CuCl_4$.

To study the photo-response characteristics of ($C_4H_9FN)_2CuCl_4$, we fabricated a prototype photodetector using a (R)-($C_4H_9FN)_2CuCl_4$ crystal. The crystal was exposed continuously to 365 nm light illumination (Figure S11c). For the fabricated photodetector, we obtained an on/off ratio ($= \frac{I_{light\ on}}{I_{dark}}$, where $I_{dark}$ is the current measured under dark environment with no light exposure) of ~1.55 (at detector bias +20 V). Although the on/off ratio obtained here for (R)-($C_4H_9FN)_2CuCl_4$ is relatively low, it should be noted that the on/off ratio for photodetectors is highly dependent on the detector bias, light illumination intensity, and wavelength of the light used for illumination. We also noticed that the photocurrent decreases over the short measurement period. This decrease is potentially due to the ionic conductivity of metal halides, which has been commonly observed in TlBr, $MAPbI_{3-x}Cl_x$, and $CsPbCl_3$ semiconductors.[13–15]

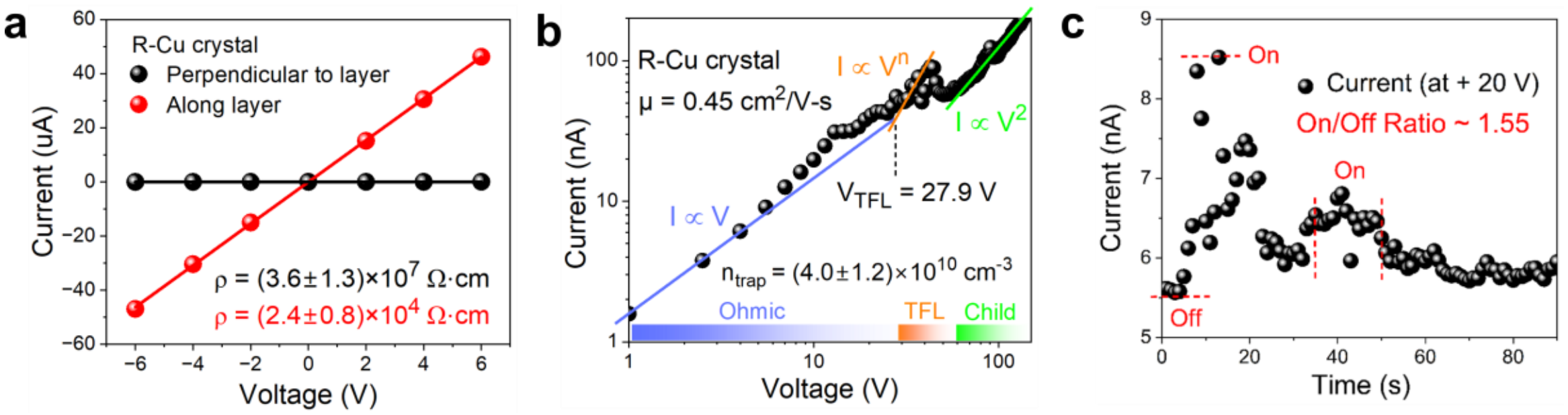

**Figure S11. (a)** I-V measurement using (R)-$(C_4H_9FN)_2CuCl_4$ crystal to determine the resistivity; **(b)** SCLC measurement data for an as-grown (R)-$(C_4H_9FN)_2CuCl_4$ crystal; **(c)** Response of (R)-$(C_4H_9FN)_2CuCl_4$ crystal to 365 nm light illumination. The fabricated photodetector was biased at +20 V.

## 10. Output from PLATON's CALC ADDSYM EXACT SHELX Command on (R)-$(C_4H_9FN)_2CuCl_4$

```
:: Current Graphics Settings: DISPLAY  = on ; DISPLAY  TYPE =  X11
:: ===(See HELP GRAPHICS)===: METAFILE = off; METAFILE TYPE =   PS

:: Data from: r-cu-300k_a.cif

:: Restricted CIF-File Format assumed (Automatic NOMOVE effective)

:: Default Global AUDIT TOOL Name: Olex2 1.5

:: Number of Valid CIF data_ ENTRIES =     1

:: Loaded Data Set: r-cu-300k

:: Extracted from CIF: r-cu-300k_a_sx.ins
:: Extracted from CIF: r-cu-300k_a_sx.hkl

::                         *** S.e.l.e.c.t.e.d  I.n.s.t.r.u.c.t.i.o.n.s ***
:: ************************ CALC ALL   - for an exhaustive geometry calculation
:: *        PLATON        * ORTEP      - for default labeled ORTEP-look-alike
:: *        ======        * VALID      - for a PLATON/checkCIF validation run
:: *    A Multipurpose    * LEPAGE     - to  check for higher metrical symmetry
:: *   Crystallographic   * ADDSYM     - for a check for MISsed SYMmetry
:: *         Tool         * NONSYM     - for a non-cryst. symm. check
:: *          --          * SOLV       - to  search for missed solvent areas
:: *(C) 1980-2026 A.L.Spek* SQUEEZE    - to  handle disordered solvents
:: *          --          * NEWMAN     - for NEWMAN-Projection Plots
:: *   version :  30426   * LIST RADII - for current atomic radii list
:: *   scratch :  400MB   * HELP       - for Available Instruction Information
:: ************************ PLUTON     - to  enter the PLUTON sub-program
::
>>INCLUDE CU CL
>>CALC ADDSYM EXACT SHELX 0.2 0.2 0.2 0.2

:: TITL r-cu-300k       P 21 21 21   R = 0.04
:: LAMBDA   0.71073
:: CELL     7.7559    8.7973    21.1758    90.000    90.000    90.000    1444.8
:: SPGR P 21 21 21    Chiral

- ADDSYM Search on ALL NON-H Atom Types [Max Allowed NonFit  0 % ]
- The  ADDSYM Search may be rerun for a choosen atom type
- Use LIST RADII for an overview of the atom types

- Number of Input Atoms Included in Search    5 (Unitcell   20)
```

```
- Density based on Input Atom Set = 1.773 g.cm-3 - Vol / Non-H atom = 21.2 Ang^3
- The Structure Implies the Following Symmetry Elements Subject to the Criteria:
- Criteria 0.20 Deg (Metric), 0.20 Ang (Rot), 0.20 Ang (Inv), 0.20 Ang (Transl)
- Pivot Atom Cu1

Symm.  Input  Reduced  (Ang)        (Deg)  %   AvrDev.(Ang)          Input Cell
Elem Cell_Row Cell_Row   d  Typ Dot Angle Fit  MaxDev.             x     y     z
---------------------------------------------------------------------------------
 c * [ 1 0 0] [ 1 0 0]  7.76  2  1     0  100   0.113  Through 0.494     0     0
                                    Cl3  -Cl1   0.193  Glide       0     0   1/2
 a * [ 0 1 0] [ 0-1 0]  8.80  2  1     0  100   0.093  Through     0 0.491     0
                                    Cl1  -Cl3   0.193  Glide     1/2     0     0
 b * [ 0 0 1] [ 0 0 1] 21.18  2  1     0  100   0.093  Through     0     0 0.498
                                    Cl1  -Cl3   0.193  Glide       0   1/2     0
-1 * ===================================  100   0.093  at      0.244 0.241 0.248
                                    Cl1  -Cl3   0.193

         T.R.A.N.S.F.O.R.M.A.T.I.O.N  M.A.T.R.I.X for CELL and HKL DATA
         ==============================================================
   Reduced->Convent         Input->Reduced       T = Input->Convent:    a' = T a
---------------------------------------------------------------------------------
(     0    -1     0 )   (     1     0     0 )   (     0     1     0 )     Det(T)
(     1     0     0 ) X (     0    -1     0 ) = (     1     0     0 )       =
(     0     0     1 )   (     0     0    -1 )   (     0     0    -1 )     1.000

Cell Lattice  a       b       c    Alpha   Beta  Gamma Volume CrystalSystem Laue
---------------------------------------------------------------------------------
Input   oP  7.756   8.797  21.176  90.00  90.00  90.00   1445 orthorhombic   mmm
Reduced  P  7.756   8.797  21.176  90.00  90.00  90.00   1445
Convent oP  8.797   7.756  21.176  90.00  90.00  90.00   1445 orthorhombic   mmm

          Conventional, New or Pseudo Symmetry
=================================================================================

Space Group  H-M:  Pbca                                     Laue:  mmm
Space Group Hall: -P 2ac 2ab                           [Schoenflies: D2h^15]
Lattice Type: oP,  Centric, Orthorhombic, Multiplicity:   8( 4), No:  61

Non-Sohncke - No Absolute Structure

  Nr            ***** Symmetry Operation(s) *****

   1               X ,              Y ,              Z
   2         1/2 - X ,            - Y ,        1/2 + Z
   3         1/2 + X ,        1/2 - Y ,            - Z
   4             - X ,        1/2 + Y ,        1/2 - Z
   5             - X ,            - Y ,            - Z
   6         1/2 + X ,              Y ,        1/2 - Z
   7         1/2 - X ,        1/2 + Y ,              Z
   8               X ,        1/2 - Y ,        1/2 + Z

:: Origin Shifted to: 0.2410, 0.2444,-0.2484 after Transformation

::                      (  0.0000  1.0000  0.0000) ( -0.2410)
:: R/t for Coordinates  (  1.0000  0.0000  0.0000) ( -0.2444)
::                      (  0.0000  0.0000 -1.0000) (  0.2484)

:: - Symmetry Elements Preceded by an Asterisk are New and Indicate
::   Missed/Pseudosymmetry.
:: - Proposed Inversion or (Glide) Planes do NOT Apply
::   for Chiral Molecules.
:: - Glide Plane Codes are with Reference to the Input Cell !!

P! r-cu-300kP 21 21 21 oP=>oP 0.3 0.000 0.00 0.193 1.00   100% Pbca

:: An SPF-style file is written to be used for the cell transformation.

:: Rounding set to OFF
```

```
:: Create r-cu-300k_a_pl.spf
:: Resd  1, SOF 1.000, Z  4, Cl4 Cu

:: Moiety_Formula = Cl4 Cu
:: Sum_Formula    = Cl4 Cu
:: Formula_Weight =      205.35 [Note: Based on SHELXL2019 Atomic Weights]
:: Formula_Z      =           4
:: SpaceGroup_Z   =           4
:: Formula_Z'     =       1.000
:: mu(MoKa)       =       21.85 cm-1 =   2.185 mm-1
:: Res.Scat.      =         409 x 0.0001
:: Friedif_stat   =         474
:: Predicted Vol  =       520.4[     520.5] Ang**3, 298[300]K

:: Resd  MolVol     Formula          - (Kitaigorodskii Analytical Formula)
!K    1    94.6 A^3 Cl4 Cu

================================================================================
Summary and Remarks : N = NOTE, W = WARNING, E = ERROR
================================================================================
W: NOMOVE option used.
:: >>> WARNING: 'CONNECTED INPUT SET' is assumed to be TRUE
:: >>> The Network Analysis may be INCORRECT when this  assumption is FALSE
N: ADDSYM finds additional (pseudo)symmetry in the structure (please check!)
N: No-Hydrogen atoms in this structure
================================================================================

:: Input Xtal Data from File r-cu-300k_a.spf - Data Type SPF14

:: NORMAL END of PLATON :     8 Pages on FILE r-cu-300k_a.lis

:: Loaded Data Set: r-cu-300k        P 21 21 21   R = 0.04 New:Pbca

:: Transformation for x,y,z coordinates
:: (  0.0000  1.0000  0.0000) (x)     -0.2410
:: (  1.0000  0.0000  0.0000) (y) +   -0.2444
:: (  0.0000  0.0000 -1.0000) (z)      0.2484

:: PAR( 22) old value =    0.3000 new =    0.7000

:: PLA100-CELL-IN     7.7559    8.7973   21.1758   90.0000   90.0000   90.0000
:: PLA100-CELL-OUT    8.7973    7.7559   21.1758   90.0000   90.0000   90.0000

:: ATOM CL3     DELETED from INPUT STREAM,  0.03 Ang. From Cl1
:: ATOM CL3     at 0.03 Ang. from Cl1     - New Av Pos:  0.0019  0.4526 -0.1054
:: ATOM CL4     DELETED from INPUT STREAM,  0.04 Ang. From Cl2
:: ATOM CL4     at 0.04 Ang. from Cl2     - New Av Pos:  0.1627  0.7285 -0.0136
:: SHELXL HKLF Matrix:  0.00  1.00  0.00  1.00  0.00  0.00  0.00  0.00 -1.00

:: Coord. Cell:     8.797     7.756    21.176     90.00     90.00     90.00
:: Refln. Cell:     7.756     8.797    21.176     90.00     90.00     90.00

:: IPR(209) old value =         0 new =         1
:: Rounding set to OFF

:: TITL r-cu-300k        P 21 21 21   R = 0.04 New:Pbca
:: LAMBDA   0.71073
:: CELL     8.7973    7.7559   21.1758    90.000    90.000    90.000    1444.8
:: SPGR P b c a
:: Resd  1, SOF 1.000, Z  4, Cl4 Cu

:: Moiety_Formula = Cl4 Cu
:: Sum_Formula    = Cl4 Cu
:: Formula_Weight =      205.35 [Note: Based on SHELXL2019 Atomic Weights]
:: Formula_Z      =           4
:: SpaceGroup_Z   =           8
:: Formula_Z'     =       0.500
:: mu(MoKa)       =       21.85 cm-1 =   2.185 mm-1
:: Predicted Vol  =       520.4[     520.5] Ang**3, 298[300]K
```

```
Space Group  H-M:  Pbca                                    Laue:  mmm
Space Group Hall: -P 2ac 2ab                      [Schoenflies: D2h^15]
Lattice Type: oP,  Centric, Orthorhombic, Multiplicity:   8( 4), No:  61

Non-Sohncke - No Absolute Structure

  Nr           ***** Symmetry Operation(s) *****

   1              X ,              Y ,              Z
   2        1/2 - X ,            - Y ,        1/2 + Z
   3        1/2 + X ,        1/2 - Y ,            - Z
   4            - X ,        1/2 + Y ,        1/2 - Z
   5            - X ,            - Y ,            - Z
   6        1/2 + X ,              Y ,        1/2 - Z
   7        1/2 - X ,        1/2 + Y ,              Z
   8              X ,        1/2 - Y ,        1/2 + Z
SFAC  C   2.31000  20.84392   1.02000  10.20751   1.58860   0.56870   0.86500 =
         51.65125   0.21560   0.00330   0.00160    11.500   0.68000  12.01000
SFAC  H   0.49300  10.51091   0.32291  26.12573   0.14019   3.14236   0.04081 =
         57.79977   0.00304   0.00000   0.00000     0.624   0.35000   1.00800
SFAC Cl  11.46041   0.01040   7.19641   1.16620   6.25561  18.51942   1.64550 =
         47.77846  -9.55741   0.14840   0.15850   678.000   0.99000  35.45000
SFAC Cu  13.33801   3.58280   7.16761   0.24700   5.61581  11.39661   1.67350 =
         64.81267   1.19100   0.32010   1.26510  5180.000   1.52000  63.55000
SFAC  F   3.53920  10.28251   2.64120   4.29440   1.51700   0.26150   1.02430 =
         26.14763   0.27760   0.01710   0.01030    51.500   0.64000  19.00000
SFAC  N  12.21261   0.00570   3.13220   9.89331   2.01250  28.99754   1.16630 =
          0.58260 -11.52901   0.00610   0.00330    19.600   0.68000  14.01000

:: Resd  MolVol     Formula          - (Kitaigorodskii Analytical Formula)
!K    1    94.6 A^3 Cl4 Cu

================================================================================
Summary and Remarks : N = NOTE, W = WARNING, E = ERROR
================================================================================
W: NOMOVE option used.
:: >>> WARNING: 'CONNECTED INPUT SET' is assumed to be TRUE
:: >>> The Network Analysis may be INCORRECT when this  assumption is FALSE
N: ADDSYM finds additional (pseudo)symmetry in the structure (please check!)
N: No-Hydrogen atoms in this structure
W: Low density (check!) of ....................................   0.944 gcm-3
N: Input Data following TRNS [e.g. (CELL/CESD)/SPGR/COORDS/UIJ)] have been
    transformed according to the specified Cell Transformation Matrix:
                       0.00000   1.00000   0.00000
                       1.00000   0.00000   0.00000
                       0.00000   0.00000  -1.00000
N: Input data (COORDINATES) have been transformed
    according to additive coordinate shift vector:
                      -0.24104  -0.24444   0.24842
N: Number of deleted ATOMS from input stream ............................    2
N: Number of Ignored Lines on INPUT .....................................    6
          of which blank in column 1 ....................................    0
================================================================================

:: Input Xtal Data from File r-cu-300k_a.spf - Data Type SPF

:: NORMAL END of PLATON :    16 Pages on FILE r-cu-300k_a.lis

:: SHELXL   res on :r-cu-300k_a_pl.res
```

## Supporting Information References